\documentclass[]{aastex63}

\newcommand{\nuvu}{N\"uv\"u}

\received{February 26, 2020}
\revised{June 12, 2020}
\accepted{June 30, 2020}
\submitjournal{ApJ}

\shorttitle{FIREBall-2}
\shortauthors{Hamden et al.}

\graphicspath{{./}{figures/}}

\begin{document}

\title{FIREBall-2: The Faint Intergalactic Medium Redshifted Emission Balloon Telescope}

\correspondingauthor{Erika Hamden}
\email{hamden@email.arizona.edu}

\author{Erika Hamden}
\affiliation{University of Arizona, Steward Observatory, 933 N Cherry Ave, Tucson, AZ 85721, USA}

\author{D. Christopher Martin}
\affiliation{California Institute of Technology, Division of Physics, Math, and Astronomy, 1200 E California Blvd, MC 278-17, Pasadena, CA 91105, USA}

\author{Bruno Milliard}
\affiliation{Laboratoire d'Astrophysique de Marseille, 38 Rue Frédéric Joliot Curie, 13013 Marseille, France}

\author{David Schiminovich}
\affiliation{Columbia University, 550 W 120th St, New York, NY 10027, USA}

\author{Shouleh Nikzad}
\affiliation{Jet Propulsion Laboratory, California Institute of Technology, 4800 Oak Grove Drive, Pasadena, CA 91109, USA}

\author{Jean Evrard}
\affiliation{Centre national d'études spatiales, 18 Avenue Edouard Belin, 31400 Toulouse, France}

\author{Gillian Kyne}
\affiliation{Jet Propulsion Laboratory, California Institute of Technology, 4800 Oak Grove Drive, Pasadena, CA 91109, USA}

\author{Robert Grange}
\affiliation{Laboratoire d'Astrophysique de Marseille, 38 Rue Frédéric Joliot Curie, 13013 Marseille, France}

\author{Johan Montel}
\affiliation{Centre national d'études spatiales, 18 Avenue Edouard Belin, 31400 Toulouse, France}

\author{Etienne Pirot}
\affiliation{Centre national d'études spatiales, 18 Avenue Edouard Belin, 31400 Toulouse, France}

\author{Keri Hoadley}
\affiliation{California Institute of Technology, Division of Physics, Math, and Astronomy, 1200 E California Blvd, MC 278-17, Pasadena, CA 91105, USA}

\author{Donal O'Sullivan}
\affiliation{California Institute of Technology, Division of Physics, Math, and Astronomy, 1200 E California Blvd, MC 278-17, Pasadena, CA 91105, USA}

\author{Nicole Melso}
\affiliation{Columbia University, 550 W 120th St, New York, NY 10027, USA}

\author{Vincent Picouet}
\affiliation{Laboratoire d'Astrophysique de Marseille, 38 Rue Frédéric Joliot Curie, 13013 Marseille, France}

\author{Didier Vibert}
\affiliation{Laboratoire d'Astrophysique de Marseille, 38 Rue Frédéric Joliot Curie, 13013 Marseille, France}

\author{Philippe Balard}
\affiliation{Laboratoire d'Astrophysique de Marseille, 38 Rue Frédéric Joliot Curie, 13013 Marseille, France}

\author{Patrick Blanchard}
\affiliation{Laboratoire d'Astrophysique de Marseille, 38 Rue Frédéric Joliot Curie, 13013 Marseille, France}

\author{Marty Crabill}
\affiliation{California Institute of Technology, Division of Physics, Math, and Astronomy, 1200 E California Blvd, MC 278-17, Pasadena, CA 91105, USA}

\author{Sandrine Pascal}
\affiliation{French Alternative Energies and Atomic Energy Commission, France}

\author{Frederi Mirc}
\affiliation{Centre national d'études spatiales, 18 Avenue Edouard Belin, 31400 Toulouse, France}

\author{Nicolas Bray}
\affiliation{Centre national d'études spatiales, 18 Avenue Edouard Belin, 31400 Toulouse, France}

\author{April Jewell}
\affiliation{Jet Propulsion Laboratory, California Institute of Technology, 4800 Oak Grove Drive, Pasadena, CA 91109, USA}

\author{Julia Blue Bird}
\affiliation{Columbia University, 550 W 120th St, New York, NY 10027, USA}

\author{Jose Zorilla}
\affiliation{Columbia University, 550 W 120th St, New York, NY 10027, USA}

\author{Hwei Ru Ong}
\affiliation{Columbia University, 550 W 120th St, New York, NY 10027, USA}

\author{Mateusz Matuszewski}
\affiliation{California Institute of Technology, Division of Physics, Math, and Astronomy, 1200 E California Blvd, MC 278-17, Pasadena, CA 91105, USA}

\author{Nicole Lingner}
\affiliation{California Institute of Technology, Division of Physics, Math, and Astronomy, 1200 E California Blvd, MC 278-17, Pasadena, CA 91105, USA}

\author{Ramona Augustin}
\affiliation{Space Telescope Science Institute, 3700 San Martin Dr, Baltimore, MD 21218}

\author{Michele Limon}
\affiliation{Department of Physics and Astronomy, University of Pennsylvania, 209 South 33rd Street, Philadelphia, PA 19104}

\author{Albert Gomes}
\affiliation{Centre national d'études spatiales, 18 Avenue Edouard Belin, 31400 Toulouse, France}

\author{Pierre Tapie}
\affiliation{Centre national d'études spatiales, 18 Avenue Edouard Belin, 31400 Toulouse, France}

\author{Xavier Soors}
\affiliation{Centre national d'études spatiales, 18 Avenue Edouard Belin, 31400 Toulouse, France}

\author{Isabelle Zenone}
\affiliation{Centre national d'études spatiales, 18 Avenue Edouard Belin, 31400 Toulouse, France}

\author{Muriel Saccoccio}
\affiliation{Centre national d'études spatiales, 18 Avenue Edouard Belin, 31400 Toulouse, France}

\begin{abstract}

The Faint Intergalactic Medium Redshifted Emission Balloon (FIREBall) is a mission designed to observe faint emission from the circumgalactic medium of moderate redshift (z $\sim$ 0.7) galaxies for the first time. FIREBall observes a component of galaxies that plays a key role in how galaxies form and evolve, likely contains a significant amount of baryons, and has only recently been observed at higher redshifts in the visible. Here we report on the 2018 flight of the FIREBall-2 Balloon telescope, which occurred on September 22nd, 2018 from Fort Sumner, New Mexico. The flight was the culmination of a complete redesign of the spectrograph from the original FIREBall fiber-fed IFU to a wide-field multi-object spectrograph. The flight was terminated early due to a hole in the balloon, and our original science objectives were not achieved. The overall sensitivity of the instrument and telescope was 90,000 LU, due primarily to increased noise from stray light. We discuss the design of the FIREBall-2 spectrograph, modifications from the original FIREBall payload, and provide an overview of the performance of all systems. We were able to successfully flight test a new pointing control system, a UV-optimized, \added{delta-doped and coated} EMCCD, and an aspheric grating. The FIREBall-2 team is rebuilding the payload for another flight attempt in the Fall of \deleted{2020}\added{2021, delayed from 2020 due to COVID-19}.

\end{abstract}
\keywords{Circumgalactic Medium, UV Spectroscopy, Multi-Object Spectroscopy, High Altitude Ballooning, UV Telescope}

\section{Introduction}\label{sec:intro}

We have built and successfully flown the Faint Intergalactic-medium Redshifted Emission Balloon (FIREBall-2), a joint mission funded by NASA and CNES. FIREBall-2 is designed to discover and map faint emission from the circumgalactic medium of moderate redshift galaxies, in particular via Ly-$\alpha$ (1216\AA), OVI (1033\AA) and CIV (1549\AA) redshifted into
the 1950-2250 \AA\ stratospheric balloon window at redshifts of z(Ly-$\alpha$)~0.7, z(OVI)~1.0, and z(CIV)~0.3. The FIREBall-2 balloon payload is a modification of FIREBall (FB-1), a path-finding mission built by our team with two successful flights (2007 Engineering, 2009 Science \citealt{2008Tuttle,2010Milliard}). FB-1 provided the strongest constraints on intergalactic and circumgalactic (IGM, CGM) emission available from any instrument at the time \citep{2010Milliard}. In contrast, FIREBall-2 was launched in a time of great strides in CGM science, with many detections of large Ly-$\alpha$ emitting structures surrounding high redshift quasars (QSOs, \citealt{2015Martin,2014Cantalupo,2016Borisova}). Despite these new, exciting discoveries, the lower redshift (z$<$2) CGM universe remains unexplored due to the inaccessibility of Ly-$\alpha$ from the ground below a redshift of 1.9 ($\sim$350 nm). The FIREBall-2 mission is currently the only telescope and instrument designed to observe this crucial component at a lower redshift.

FIREBall-2 consists of:
\begin{enumerate}
\item A 1-meter primary parabolic mirror
\item A sophisticated 6 axis attitude and pointing control system, yielding less than 1 arcsec RMS over long integration times.
\item A delta-doped, AR-coated UV optimized electron multiplying CCD (EMCCD), which provided $>$50\% efficiency in the 1980-2130 \AA\ bandpass.
\item The first balloon flight of a multi-object spectrograph, using pre-selected targets and custom spherical slit masks.
\item A flight test of an aspheric anamorphic grating developed at HORIBA Jobin Yvon. The grating acted as a field corrector.
\item Partnership between NASA and CNES. \deleted{The team is shown in Figure \ref{fig:team}.}
\item 4 completed PhDs (including 3 women), with an additional 7 students receiving significant mission training as part of their thesis work.
\end{enumerate}

The 2018 flight of FIREBall-2 occurred on September 22nd from Fort Sumner, NM, with launch support provided by NASA's Columbia Scientific Ballooning Facility (CSBF). During the flight, all systems of this sophisticated payload performed as expected. The balloon and payload reached an altitude of 39 km several hours after launch and then began a slow descent as a result of a hole in the 40 million cubic foot (MCF) balloon. The balloon flight was terminated after 4 hrs of dark time, with less than one hour spent above the minimum science altitude of 32 km. Upon landing, the payload suffered some structural damage that the team is working to repair. We expect a second launch of the refurbished payload in \deleted{2020}\added{2021, due to delays from COVID-19}.

\added{This paper presents an overview of the FIREBall-2 mission. In particular, we describe changes from the previous version of the mission, FIREBall-1, and both the motivation and results of those changes. We provide performance details about all subsystems, and references to more specific papers about those subsystems in the case of the detector and CNES guidance systems. We give an overview of the analysis of the data collected during flight, which was severely limited by the low altitude and high background.}

\added{The paper is arranged as follows. We briefly describe the current state of the art of CGM science in Section \ref{sec:science}. Previous flights of the FIREBall payload are described in Section \ref{sec:previous}, with changes from the earlier spectrograph design detailed in Section \ref{sec:redesign}. Major instrument components and their performance are described in Section \ref{sec:components}, including the telescope (Section \ref{sec:telescope}), spectrograph optical design (Section \ref{sec:spectrodesign}), delta-doped UV detector (Section \ref{sec:emccd}), cooling and vacuum system (Section \ref{sec:cooling}), thermal control system (Section \ref{sec:thermal}), coarse and fine guidance systems (Sections \ref{sec:pointing} and \ref{sec:guider}), and the communications system (Section \ref{sec:comms}). The flight itself is described in Section \ref{sec:Flight}, including anomalies (Section \ref{sec:balloon}), target fields observed (Section \ref{sec:observations}), and overall in flight performance (Section \ref{sec:performance}). Some preliminary data analysis is shown in Section \ref{sec:data}, covering noise from smearing, cosmic rays, scattered light, and overall sensitivity. We discuss the future of the FIREBall mission in Section \ref{sec:future} and the importance of continuing UV CGM science in future, more ambitious telescopes in Section \ref{sec:space}.}

\deleted{[comment=Deleted Figure of FIREBall-2 Team. Some issues with formatting figure deletions.]}

\subsection{State of the art in CGM science}\label{sec:science}

Studies of the circumgalactic medium (CGM) are now entering their second decade. Since the installation of the Cosmic Origins Spectrograph (COS) on HST \citep{2012Green}, observations of the CGM were conducted via absorption line studies of foreground galaxies using background QSOs \citep{Croft+02,2013Tumlinson,Lee+14,2014Bordoloi}, coincident quasars (Quasars Probing Quasars, \citealt{2006Hennawi}), and blind searches for galaxies around very bright quasars (Keck Baryonic Structure Survey, \citealt{2012Rudie}). These observations provided the first view of the CGM around nearby galaxies, but were limited by the method.  

These surveys have conclusively detected absorption from HI, MgII, CIV, OVI and other ionized metals within the projected virial radius of foreground galaxies. Because these studies rely on serendipitous sightlines that typically illuminate only a single beam through the CGM of a galaxy, they are powerful but limited. There is no way to distinguish between the presence of gas in filaments, a dispersed collection of gas clumps, or a smooth distribution of gas within the CGM. Statistical stacking of these galaxies produces a composite picture of the CGM, but at the price of smoothing out revealing spatial/kinematic structures, including disks, bipolar outflows, warps, tidal tails, filaments, etc.  Any one galaxy typically does not have multiple background QSOs and so most studies have used statistical combinations of similar type galaxies to develop an overall picture of the CGM. These studies have pointed the way forward for more targeted, but significantly more challenging, observations of direct emission from the CGM.

Improved detectors, higher throughput optics, and state-of-the-art instruments, like the Palomar and Keck Cosmic Web Imagers (PCWI \& KCWI, \citealt{2014Martina, 2014Martinb, 2015Martin, 2016Martin}) and Multi Unit Spectroscopic Explorer (MUSE, \citealt{2016Borisova, AB+16, AB+19}), are pioneering a new field of IGM astronomy at high redshifts where Ly-$\alpha$ is observable at visible wavelengths.

After the first exciting discoveries of the Slug and other similar nebulae \citep{2014Cantalupo,2015Martin}, there has been an increasing body of work examining this new population of CGM structures. \citet{2016Borisova} have found a 100\% incidence of large Ly-$\alpha$ nebula around z$>$3 QSOs. \citet{2016Wisotzki} observed the Hubble Deep Field South, finding extended Ly-$\alpha$ halos around 21 out of 26 galaxies with a redshift range of 3$<$z$<$6. \citet{2018Cai} found one of many extremely large Ly-$\alpha$ nebula as part of a close QSO pair (around z$\sim$ 2.3), linking cosmic over-densities to these structures. A follow up paper in 2019 found 16 QSOs with extended emission at lower redshifts (2.1$<$z$<$2.3) \citep{2019Cai}. \citet{2019Martin} has found that these nebulae have complex velocity structures that may indicate possible rotation and radial inflow. \citet{2019OSullivan} has observed 49 galaxies at 2.2$<$z$<$2.8. This survey finds the incidence of large Ly-$\alpha$ structures to be less frequent and with a lower covering fraction than those at higher redshift, likely indicating an evolutionary change between redshifts. 

The realm of CGM direct imaging is still in its infancy and there have been few attempts to link these individual observations to the larger surveys conducted via absorption lines. In addition, for the lowest redshift galaxies where star formation rates are declining throughout the universe, CGM direct imaging is not possible on the ground \added{due to the limits of atmospheric transmission}.

\added{The typical blue cutoff on the ground is around 350nm, where extinction is primarily driven by Rayleigh scattering \cite{2013Buton}. Below 320nm, the primary absorber is Ozone, which remains significant through the ultraviolet. At sufficient altitudes (above ~30 km), there is a transmission window between 195 and 220 nm where ozone and molecular oxygen both have troughs in their absorption cross sections \citep{2012Matuszewski,1971Ackerman}. The transmission increases with altitude. Figure \ref{fig:balloonwindow} (Figure 3.2 from \citep{2012Matuszewski}) shows atmospheric transmission at 34 km.

\begin{figure}[t]
    \centering
    \includegraphics[width=0.75\textwidth]{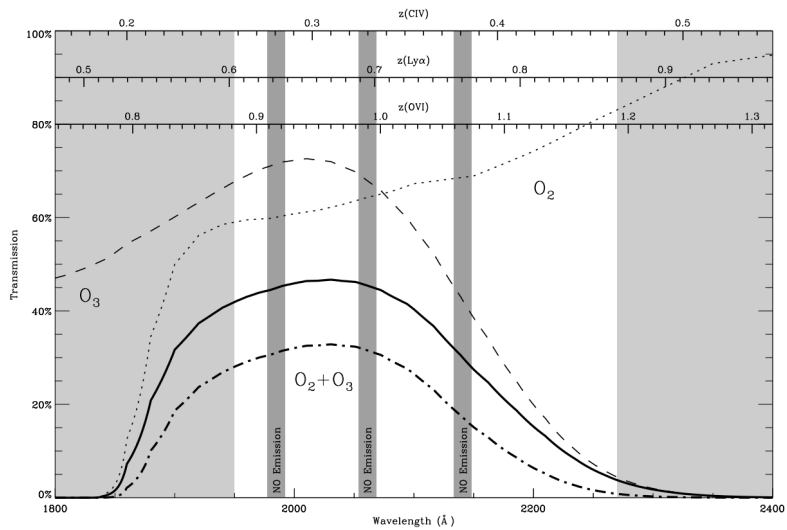}
    \caption{\emph{Figure 3.2 and caption from \citet{2012Matuszewski}}. A computed atmospheric transmission curve for observations from an altitude of 34 km (112 kft, 7 mbar), roughly that for the second FIREBall flight in 2009. The calculation included absorption by O2, N2 and N2O, and Rayleigh scattering effects. The solid curve shows the transmission for a target at the maximal elevation of the FIREBall telescope, 70$^{\circ}$, the dot-dashed curve for the minimum, 40$^{\circ}$. The dashed and dotted curves show the oxygen (O2) and ozone contributions to the transmission losses. The light gray areas lie outside of the FIREBall bandpass; the narrow bands near the center correspond to three nitric oxide airglow bands. The accessible redshift ranges for the three principal FIREBall emission lines are overplotted on the axes near the top of the image.} 
    \label{fig:balloonwindow}
\end{figure}
}

In this rich landscape, FIREBall-2 provides crucial detections of the CGM around several hundred low redshift galaxies, a unique capability at a perfect time. 

\subsection{Previous flights of FIREBall}\label{sec:previous}

The FIREBall telescope has flown two previous times, from Palestine, TX in 2007 \citep{2008Tuttle} and from Fort Sumner, NM in 2009 \citep{2010Milliard}. The original FIREBall instrument (FB-1) consisted of a fiber-fed integral field unit (IFU), which used an Offner spectrograph design and a GALEX NUV spare microchannel plate (MCP) as the detector. 

The FB-1 spectrograph had a resolution of R=5000, a plate scale of 12 $\mu$m arcsec$^{-1}$, an angular resolution of 10 arcsec, and a circular field of view with a diameter of 160 arcsec. The fiber bundle consisted of 300 fibers with a core diameter of around 8 arcsec per fiber. The overall system throughput was 0.5\% \citep{2008Tuttle}. The telescope (1 m parabolic primary mirror and 1.2 m flat siderostat mirror) and gondola structure remained unchanged between all flights of FIREBall, including FB-1 in 2007 and 2009 and the most recent FIREBall-2 flight in 2018. 

\subsubsection{2007 Flight}

The 2007 flight of FIREBall-1 was an engineering flight. The telescope was launched on July 22nd, 2007 from the CSBF location in Palestine, TX. This flight achieved 3 hours of dark time while the total time of flight was 6 hours. The instrument was only able to maintain pointing for up to 1 minute at a time due to a pivot failure that occurred during the launch and severely limited pointing control.

\subsubsection{2009 Flight}

The 2009 flight was launched from Fort Sumner, NM on June 7th, 2009. This payload used the same spectrograph as the 2007 flight, with an improved fiber bundle and a reinforced pivot. The balloon reached an average altitude of 113 kft (yielding 25\% atmospheric transmission) due to an unusual monsoon weather pattern, rather than the desired 120 kft (up to 80\% transmission). The payload performed flawlessly and obtained a full night of observations on three science targets. 

The science targets consisted of a section of the GROTH strip \citep{2007Davis}, QSO PG1718+481 \citep{2003Crighton}, and the DEEP2 ZLE field \citep{2002Simard}. Due in part to low instrument throughput, no CGM emission was detected from these targets down to a sensitivity limit of $\sim$75,000 LU for a single galaxy, or 23,000 LU for a stack of 10 (1 LU = 1 photon cm$^{-2}$ s$^{-1}$ sr$^{-1}$). This lack of detection motivated the complete redesign of the FIREBall-2 spectrograph as described in Section \ref{sec:redesign}.

\subsection{Changes between FIREBall-1 and FIREBall-2}\label{sec:redesign}

The spectrograph was redesigned to increase the chances of detecting emission from the CGM of z=0.7 galaxies via Ly-$\alpha$. The changes increased the field of view, number of targets per observation, and overall instrument throughput. In addition, galaxies were pre-selected based in part on likelihood of expected emission. 

A two-mirror field corrector was designed and added to the optical path after the prime focus to increase the usable field of view to a $\sim$30 arcmin diameter circle. The field of view was a 28x12 arcmin$^2$ rectangle set by the size of the detector. The fiber bundle was replaced with a multi-object slit mask, which could target up to 70 galaxies per field, although this density of targets causes some spectral overlap. A rotating carousel allows for selection between nine different masks with 4 designed for particular galaxy fields. Four other masks are used for calibration and one slot is left empty. Using a slit mask instead of a fiber bundle increased the sensitivity by eliminating the potential for UV absorption in the fibers. The angular resolution was also improved, from 10 arcsec to 4.5 arcsec through a new spectrograph design, in order to better separate the galaxy emission from the CGM signal. The GALEX spare NUV MCP detector was replaced with a high efficiency delta-doped EMCCD, which increased throughput by a factor of 8. 

The change in detector from an MCP to an EMCCD drove additional technical changes from FB-1. This change added a requirement that the spectrograph track the sky for long ($\sim$100 s) exposures. A MCP is able to time tag photons and therefore the instrument does not need sidereal tracking ability, while an EMCCD would be read out at longer than 100 second intervals, necessitating better pointing. The use of the slit mask also demanded an improved pointing system to keep all targets in their slits without excess jitter. The gondola pointing system was improved to provide significantly finer pointing stability (requirement went from 6 to $<$ 2 arcseconds rms error, although the in flight performance was better). The CCD detector must be operated at $<$ -100$^{\circ}$ C to reduce thermal noise, and so required a cryocooler and associated cold chain, charcoal getter, and vacuum system. The CCD controller, CMOS guider, and cryocooler reject heat also required significant cooling and the addition of a thermal control system.

\subsubsection{Overall Sensitivity Improvements}

As described in the proceeding sections, FIREBall-2 is designed to increase the sensitivity over FB-1. The expected sensitivity calculation is conducted in detail in \citet{2019Picouet}, and is briefly summarized here. The calculation for expected performance is based on throughput measurements of the optical system, noise performance of the detector, expected sky background, and nominal altitude transmission at 40k or 130 kft. The anticipated sensitivity is 8,000 LU, an improvement of a factor of 8 over FB-1. 

\section{Major instrument components}\label{sec:components}

\begin{figure}[t]
    \centering
    \includegraphics[width=0.45\textwidth]{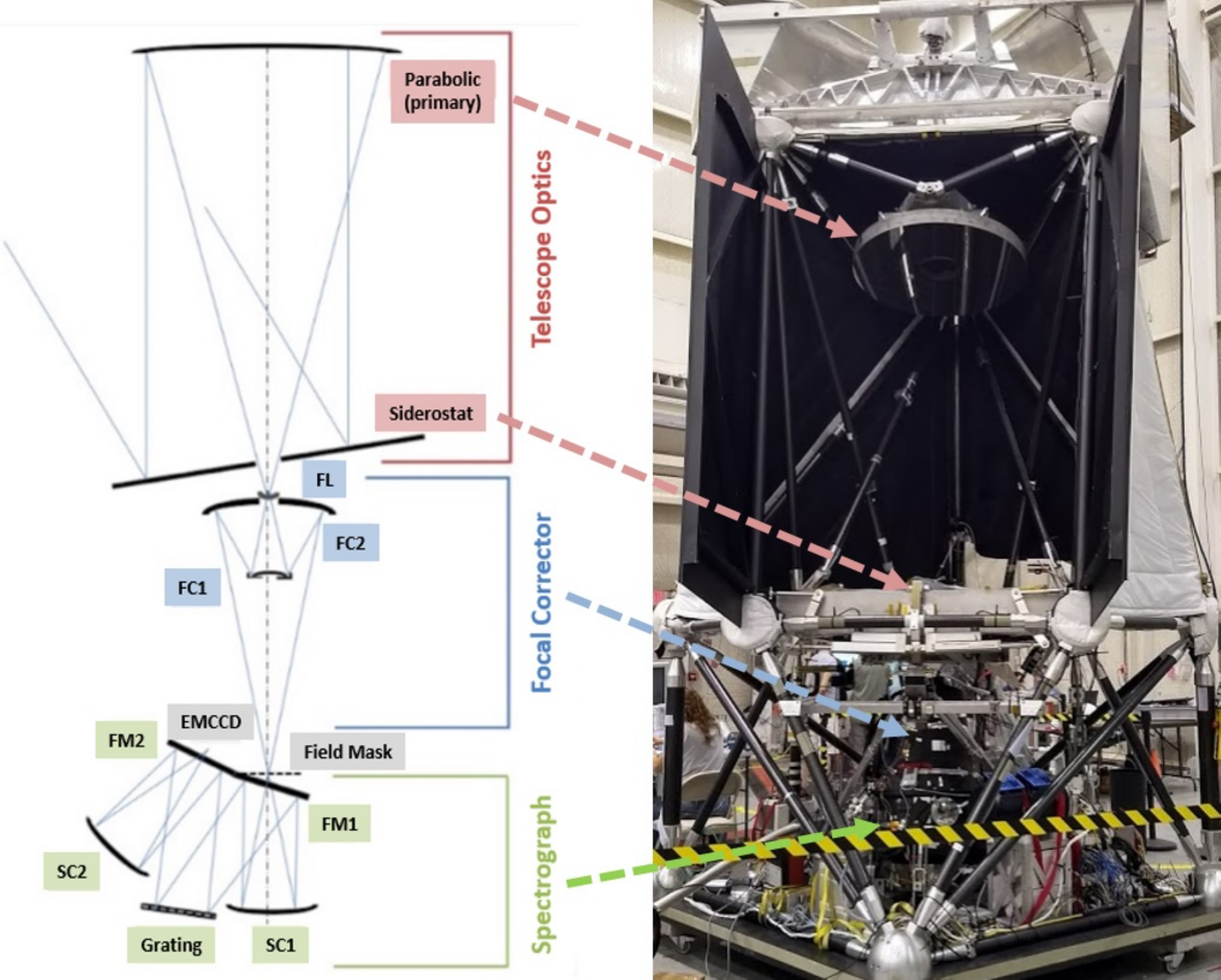}
    \caption{\emph{Left: }The FIREBall-2 light path through the entire instrument assembly. \emph{Right:} FIREBall-2 as-built, awaiting launch.} 
    \label{fig:telescope}
\end{figure}

The major instrument components of FIREBall-2 are described below. The telescope and instrument are shown in Figure \ref{fig:telescope}. More detail is provided in additional papers that focus on the fine guidance system \citep{2019Montel}, the overall attitude control, the calibration strategy \citep{2019Picouet}, and the detector performance \citep{2019Kyne}. Where appropriate, the performance in flight for each subsystem is also described. The overall instrument performance is described in Section \ref{sec:performance}.

\subsection{Telescope Assembly and Gondola}\label{sec:telescope}

The FIREBall-2 telescope and gondola are the same structures described in \citet{2008Tuttle} and \citet{2010Milliard}. The telescope assembly consists first of a flat 1.2 m siderostat mirror, which provides elevation control between 40-70$^{\circ}$ altitude and coarse x and y pointing via a tip/tilt frame. \added{the elevation limits are due to a mechanical hard stop at 40$^{\circ}$ and the balloon and top of gondola at 70$^{\circ}$. The flight train is about 90 m in this case.} The siderostat feeds the primary mirror, an f/2.5 1 m parabolic mirror. Both optics had survived two previous descents and landings. The parabola debonded during the 2007 landing, but was otherwise unharmed. The mirror coatings were stripped by Optical Mechanics, Inc., which originally fabricated them, and were then re-coated at Goddard Space Flight Center with an Al/MgF$_2$ coating optimized for 205 nm.

The gondola structure consists of carbon-fiber rods with a sparse set of connecting nodes, forming a stiff (first resonance above 20 Hz), thermally-stable kinematic structure. Optical mounts to the gondola have been athermalized to compensate for residual expansion effects in the rods and the aluminum couplings. The gondola has also flown two previous times, sustaining damage typically only to the carbon fiber rods, which are replaced between flights. 

\subsection{Instrument Optical Design}\label{sec:spectrodesign}

The choice to increase the field of view and change from a fiber-fed IFU to a multi-object mask spectrograph necessitated a complete redesign of the spectrograph optics. A comprehensive discussion of the optical design and specifications is provided in \citet{2016Grange}. Briefly, there is a two mirror field corrector to increase the quality of the field over a 30 arcmin field of view. The slit mask is a spherical surface to match the focal plane from the field corrector. The spectrograph consists of two Schmidt mirrors (one as collimator and one as camera) with folding flats for compactness and an aspherized reflective Schmidt grating. The grating was manufactured using a double replication process at HORIBA Jobin Yvon. It is a novel high throughput cost-effective holographic grating with a groove density of 2400 l mm$^{-1}$ over a 110x130 mm aspherized reflective surface \citep{2014Grange}. The shape of the grating corrects for the spherical aberrations of the rest of the optical system. The grating consists of an aluminum substrate with native oxide. Back up gratings were capped with a 70nm thick layer of MgF$_2$ to optimize in the FIREBall-2 band pass and a 28$^{\circ}$ angle of incidence, but were not used in the 2018 flight. The flight grating reflectance exceeded 50\% in the band pass, a significant improvement over the FIREBall-1 grating performance of 17\% \citep{2014Quiret}. The grating provides a slit-limited resolution of R$\sim$2000 at the detector. The as built instrument performance is discussed in greater detail in Section \ref{sec:performance} and in \citet{2019Picouet}.

\subsection{Delta-doped AR-coated EMCCD}\label{sec:emccd}

The electron multiplying CCD used on this flight was a Teledyne-e2v CCD201-20 architecture with 13 $\mu$m square pixels. Nominally the CCD201s are frame transfer devices and have a 1k $\times$ 1k image area and storage region. However, FB-2 requires use of the entire pixel array for a larger FOV and so the detector is read out in line transfer mode as a 2k $\times$ 1k device. These detectors also have an additional 1k pixel extension to the serial register. Out of these, 604 act as multiplication pixels. When these pixels are clocked with a voltage above $\sim$39 V, their wells are deep enough to allow impact ionization of electrons as they are moved through the serial register. The net result of this is that single electron events in the image area will be multiplied to many times above the read noise, significantly increasing the signal-to-noise ratio (SNR). A more detailed description of the operation of these devices can be found in \citep{2008Daigle,2010Daigle,2010Tulloch,2011Tulloch,2019Kyne}.

The flight device was the end result of several years of technology development undertaken by JPL, Caltech, and Columbia University for use in the UV, funded by NASA. \added{This development builds on JPL's pioneering work on delta-doping and UV detector development.} The maturation of these devices is detailed in a series of papers \citep{2015Jewell,2016Hamden,2017Nikzad,2019Kyne}.

\begin{figure}
\centering
    \includegraphics[width=0.45\textwidth]{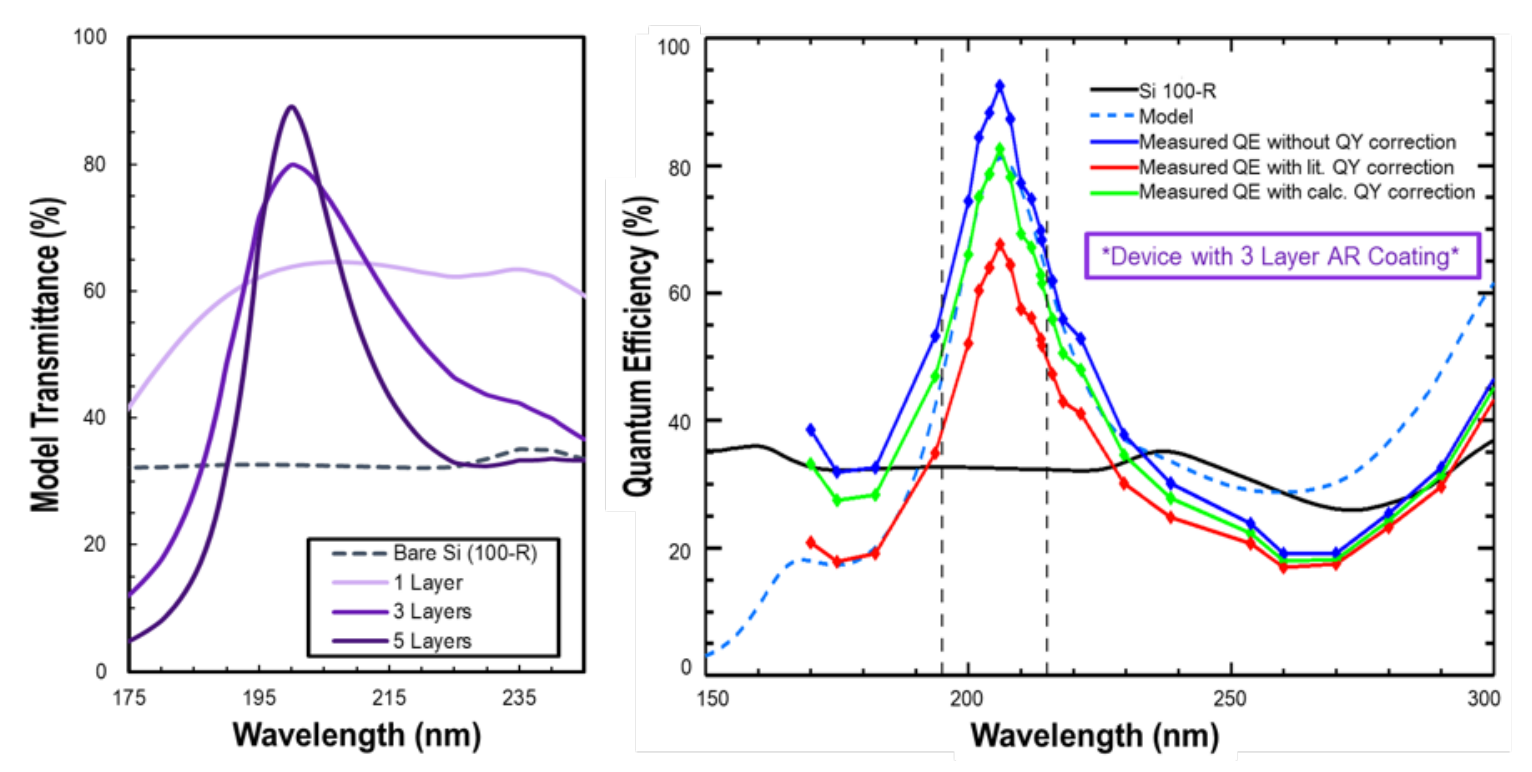}
    \caption{Left: Model transmittance/performance for 2D-doped silicon detectors with multilayer AR coatings tailored for the FIREBall bandpass. Adding complexity increases peak QE, but results in a narrower peak. Theoretical response for a bare 2D-doped silicon detector is also shown. Left: Experimental results for 2D-doped silicon detector with the 3-layer FIREBall AR coating. As measured QE is shown alongside QE corrected for quantum yield \citep{2016Hamden,1998Kuschnerus}.} 
    \label{fig:QE}
\end{figure}

The detector performance was verified via testing at JPL and Caltech, with additional validation of performance at Teledyne-e2v. JPL testing included QE verification (measurement shown in Figure \ref{fig:QE}) following the method described in \citet{2010J}, while Caltech measured QE at limited wavelengths using \nuvu\ v2 and v3 CCCP controllers and a custom flight printed circuit board (PCB). \added{Caltech also independently tested the detectors on sky at Palomar using the Cosmic Web Imager instrument \citep{2010Matuszewski}.}  These controllers provide 10 and 5 ns granularity, respectively, to optimize the pixel clocking strategy, readout speed, and wave form shape/height to minimize clock-induced-charge (CIC), read noise, and deferred charge \citep{2015Hamden,2016Kyne}. The PCB was designed using the suggested configuration from \nuvu\ to reduce additional read noise from the readout process. Total read noise in the lab camera system was 100 e$^-$. Measured CIC from lab data was 4.2 $\times$ 10$^{-3}$ e- pix$^{-1}$ frame$^{-1}$ in the serial register and 7 $\times$ 10$^{-4}$ e- pix$^{-1}$ frame$^{-1}$ in the parallel clocking, for a total of 4.9 $\times$ 10$^{-3}$ e- pix$^{-1}$ frame$^{-1}$. The dark current level was 8 $\times$ 10$^{-2}$ e- pix$^{-1}$ hr$^{-1}$ at a temperature of -115$^{\circ}$C.

The detector was installed into the spectrograph in the spring of 2016, and was extensively tested in the spectrograph system. Both the detector and \nuvu\ v2 controller performed reliably and with stability throughout integration and testing showing no change in behavior between installation and flight. A brief discussion of detector noise and performance is in Section \ref{sec:data} and is discussed in detail in \citet{2019Kyne}.

\subsection{Cooling and vacuum system}\label{sec:cooling}

\begin{figure}[]
    \centering
    \includegraphics[width=\columnwidth,trim=0in 0.0in 0in 0in]{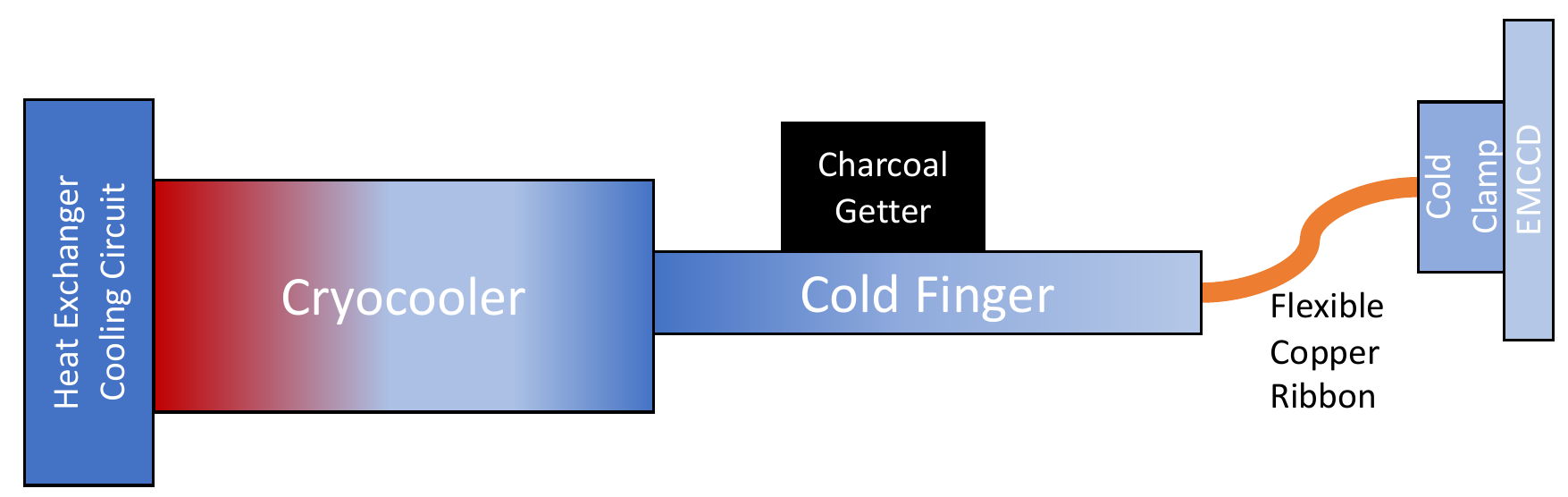}
    \caption{Schematic of detector and getter cooling system. Both are cooled by the cryocooler, and connected to the cryocooler cold head by a gold plated copper cold finger. A flexible copper ribbon connects the end of the cold finger to the cold clamp which cools the back of the EMCCD detector. The charocoal getter is directly connected to the cold finger.} 
    \label{fig:coldchain}
\end{figure}

The choice to change the detector from a MCP to a CCD required the addition of a cooling system to reduce the noise contribution from dark current. To avoid the build up of ice or other contaminants on the detector surface, and maintain an effective vacuum during the flight, a cryopumping system was developed. A schematic of the cooling system is shown in Figure \ref{fig:coldchain}.

A Sunpower CryoTel CT cryocooler was used to cool both the detector and charcoal getter. The CT provides up to 120 W of cooling power, providing a lift of greater than 10 watts at cryogenic temperatures. This was sufficient to cool an 0.75 liter charcoal getter and maintain a pressure of less than a few 10$^{-6}$ Torr for the expected $\sim$24 hour flight time. In order to achieve this vacuum, a careful strategy of tank and component bakeouts was implemented. The spectrograph tank interior (black anodized aluminum) itself was a significant source of water in the vacuum system and needed to be baked out at $>$50$^{\circ}$ C for several weeks to achieve and maintain a high vacuum.

The flight charcoal getter interfaces with a gold plated solid copper coldfinger, which has a one-inch diameter circular cross section and connects the cold head of the cryocooler to the rest of the system. There is a single joint connection in the cold finger to enable installation in a tightly packed spectrograph tank. The getter is the first thermal load on the cryocooler and can be kept at a much lower temperature than the EMCCD (typically $\sim$75$^{\circ}$ C colder than the detector). The EMCCD is held by a cold clamp coupled to the far end of the coldfinger via a flexible cold ribbon. All cold surface interfaces have a layer of indium for better contact and thermal conductivity.

In flight, the cryocooler was operated at 120 W to maintain the charcoal at a temperature of -180$^{\circ}$ C and the detector at -115$^{\circ}$ C. 

\subsection{Thermal control system}\label{sec:thermal}

An additional requirement for FIREBall-2 was removing excess heat generated by a number of systems: The cryocooler, the \nuvu\ EMCCD controller, and the pco.edge 5.5 guider camera (discussed in Section \ref{sec:guider}) all generated waste heat that needed to be managed in flight. The expected pressure at float altitude of 3 mbar presents a difficult thermal environment in which normal \deleted{radiative}\added{convective} cooling is not effective.

The cryocooler in particular was most sensitive to \added{lack of convective} cooling, as the temperature of the heat rejection point is directly coupled to the temperature of the cold head. An inability to cool the reject point would result in inefficient cooling, a higher detector operating temperature, and higher dark current. While the cooler can operate with a reject of up to 80$^{\circ}$ C, this is a significantly less efficient operating mode than at lower reject temperatures. A thermal vacuum test conducted in the winter of 2017 at CNES in Toulouse indicated the need to actively cool the cryocooler heat rejection point.

To address this we used a water circulation system. It consisted of a dewar filled with 0$^{\circ}$ C ice and water with a narrow outlet, allowing for thermal evaporation to the 3 mbar atmosphere to maintain a low temperature in the water dewar. A circulating water circuit passed through the dewar before reaching water blocks connected to the crycooler rejection point and body, the \nuvu\ pressure vessel, and thermal blocks connected via copper straps to the guider camera pressure vessel inside the spectrograph tank. The dewar volume of 20 liters had sufficient cooling power for 24 hours of operation. Both the \nuvu\ and guider camera were turned off during the ascent and daytime float phases of the flight to save cooling power for night time operations. A schematic of the cooling approach is shown in Figure \ref{fig:coolingblock}. \added{We saw no evidence of any impact of the dewar system on the guidance due to, for example, sloshing of the cooling liquid. The mass of the cooling liquid was marginal compared to the rest of the payload (0.7\% of the total mass).}

\begin{figure}[]
    \centering
    \includegraphics[width=\columnwidth]{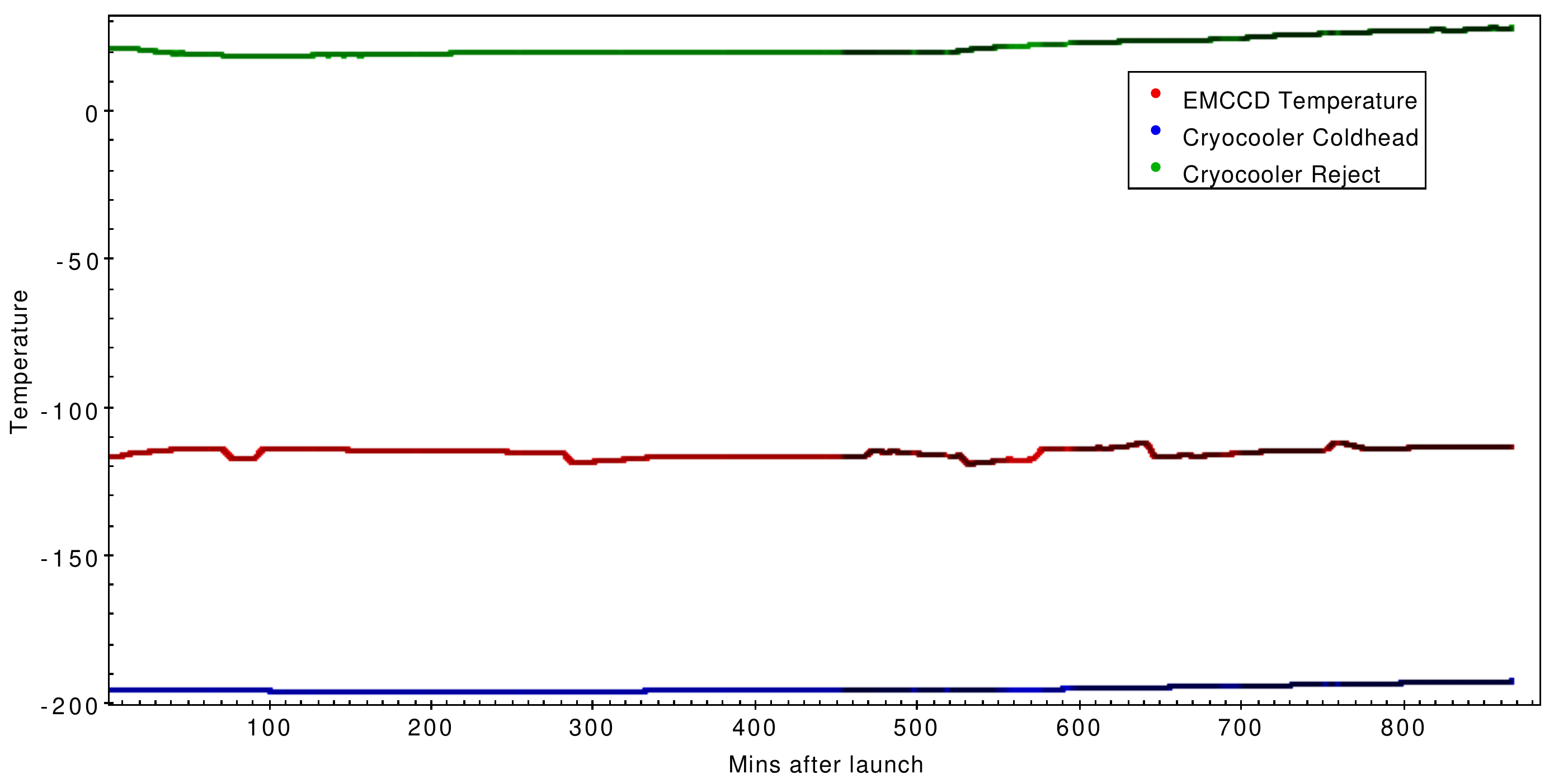}
    \caption{Figure of temperatures for three critical parts of the cooling chain: The EMCCD (red), the cyrocooler coldhead (blue), and the cryocooler reject (green). The three temperatures are correlated, with increases in the reject temperature resulting in increases in the cyrocooler cold head and therefore eventually EMCCD temperature. A heater on the EMCCD maintains a constant temperature, so the EMCCD measurement will not be immediately impacted by the reject temperature increase. In the last 200 minutes of the flight, the reject temperature started to increase due to the loss of cooling capacity at lower altitudes as described in Section \ref{sec:thermal}.}
    \label{fig:Flighttemp}
\end{figure}

\begin{figure}[]
    \centering
    \includegraphics[width=\columnwidth,trim=0in 0.5in 0in 0in]{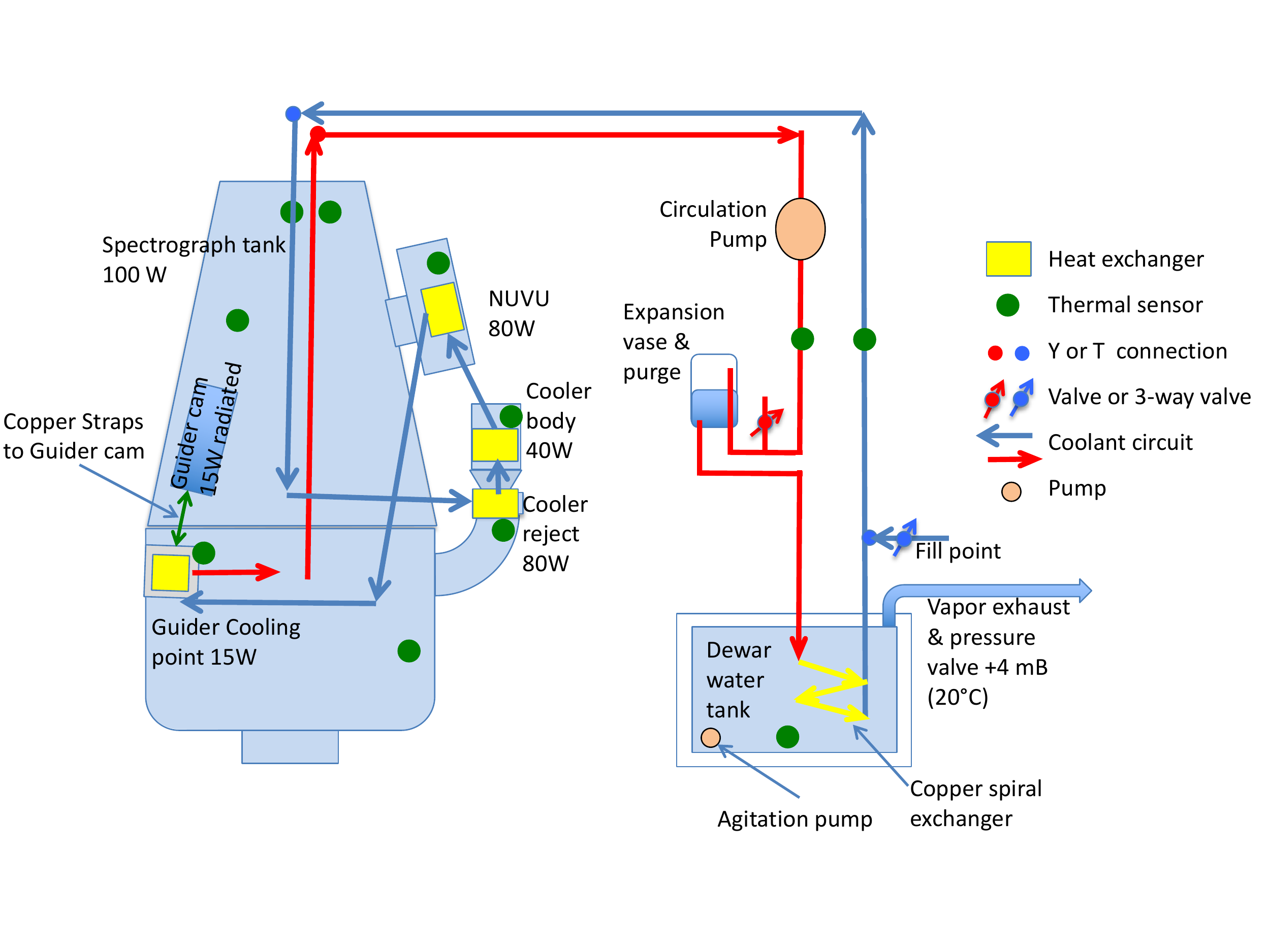}
    \caption{Block diagram of cooling scheme. Spectrograph tank and items requiring active cooling are shown as blue blocks, with blue and red lines indicating the cooling circuit. There are two identical cooling circuits, splitting and re-joining at the top of the tank. For clarity, only one is shown. The configuration shown is the same as during the flight, with a water ice solution in the dewar. During ground testing, a cooling circuit is used instead of water ice.} 
    \label{fig:coolingblock}
\end{figure}

During the flight, the cooling system maintained a cryocooler reject temperature of 19$^{\circ}$ C at 100 W of cooling power. Because of the short duration of the flight, less than one third of the water volume was consumed. However, the lower altitude experienced during the flight meant the evaporative cooling scheme was less efficient than anticipated. Towards the end of the flight the temperature of the cooling system, including cryocooler reject, cold head, charcoal getter, and detector, started to trend upwards, which can be seen in Figure \ref{fig:Flighttemp}. By this time, the payload altitude was so low that this did not significantly impact data collection and the flight was terminated shortly afterward.

\subsection{Coarse guidance system}\label{sec:pointing}

The attitude control system for FIREBall-2 was designed primarily by CNES. It is an update from the generic CNES system and the one used on the previous two flights of FIREBall. A detailed discussion of the basis for the system can be found in \citet{2019Montel}, and more detailed results from the flight will be published in an upcoming paper. The pointing requirements are extremely constrained for a balloon payload ($<$ 1" pointing stability in three axes over several hours), as the system is frequently disturbed and in motion. The 3 degrees of freedom of the instrument are controlled through a 4-axis control system which uses:
\begin{enumerate}
\item Azimuth control of the gondola from the pivot connection to the balloon flight train
\item 2-axis control of the flat siderostat mirror (tip/tilt in elevation and cross elevation)
\item A rotation stage that also serves as the spectrograph tank mount.
\end{enumerate}

The CNES attitude control system (ACS) uses multiple sensors to determine the position and actuate each of the four axes of control. The azimuth fine pointing system consists of an IMU-gyrocompass and anologic gyro (\added{I}XBLUE, France) that are used to measure azimuth. This system controls the azimuth via the pivot, which links the telescope payload to the balloon. The boresight fine pointing system uses both the IMU and fine guider (described in Section \ref{sec:guider}). This system controls elevation and cross elevation via two fine pointing actuators, encoders, and two reaction wheels. The field rotation pointing system uses the rotation error signal from the fine guider. This system controls field rotation via the spectrograph tank rotation stage and encoder. 

A large field of view attitude sensor (ASC from DTU, Denmark) was also on board as a back-up sensor in case of guider failure and for coarse positioning information. The DTU was mounted on the siderostat frame and observed a region of the sky adjacent, but not overlapping, with the science FOV. The DTU sensor is a well known star tracker that can deliver a 10" 3-axis attitude measurement. Both the DTU and IMU systems from CNES were the same as the ones used on FB-1. 

The flight train of FIREBall-2 can be modeled as a multiple torsion double pendulum. The CNES system is able to predict and account for many of the expected modes of the gondola and provide damping of the dynamic modes. There are both \deleted{slow and fast} pendulum motion modes and \deleted{slow and fast} wobbling modes. The pendulum modes are well understood\added{, corresponding to roll and pitch,} and behaved as expected during the flight. \added{The primary low frequency pendulum modes have periods of 23s and 9s, while the high frequency wobbling modes have periods of 1.8s and 2.2s.} The high frequency modes are damped with an active damping system using 2 reaction wheels. The wobbling frequencies in particular are sensitive to the ladder length below the parachute and the mass of the payload. The amplitude of the wobbling modes is directly proportional to the quality of the pointing. Changes in balloon altitude or changes in the rotator torque excited the wobbling modes and were difficult to dampen. This resulted in operating with reduced gains on the azimuth pointing control loop and adjustment of the roll reaction wheel filter frequency. 

\begin{figure}[]
    \centering
    \includegraphics[width=\columnwidth]{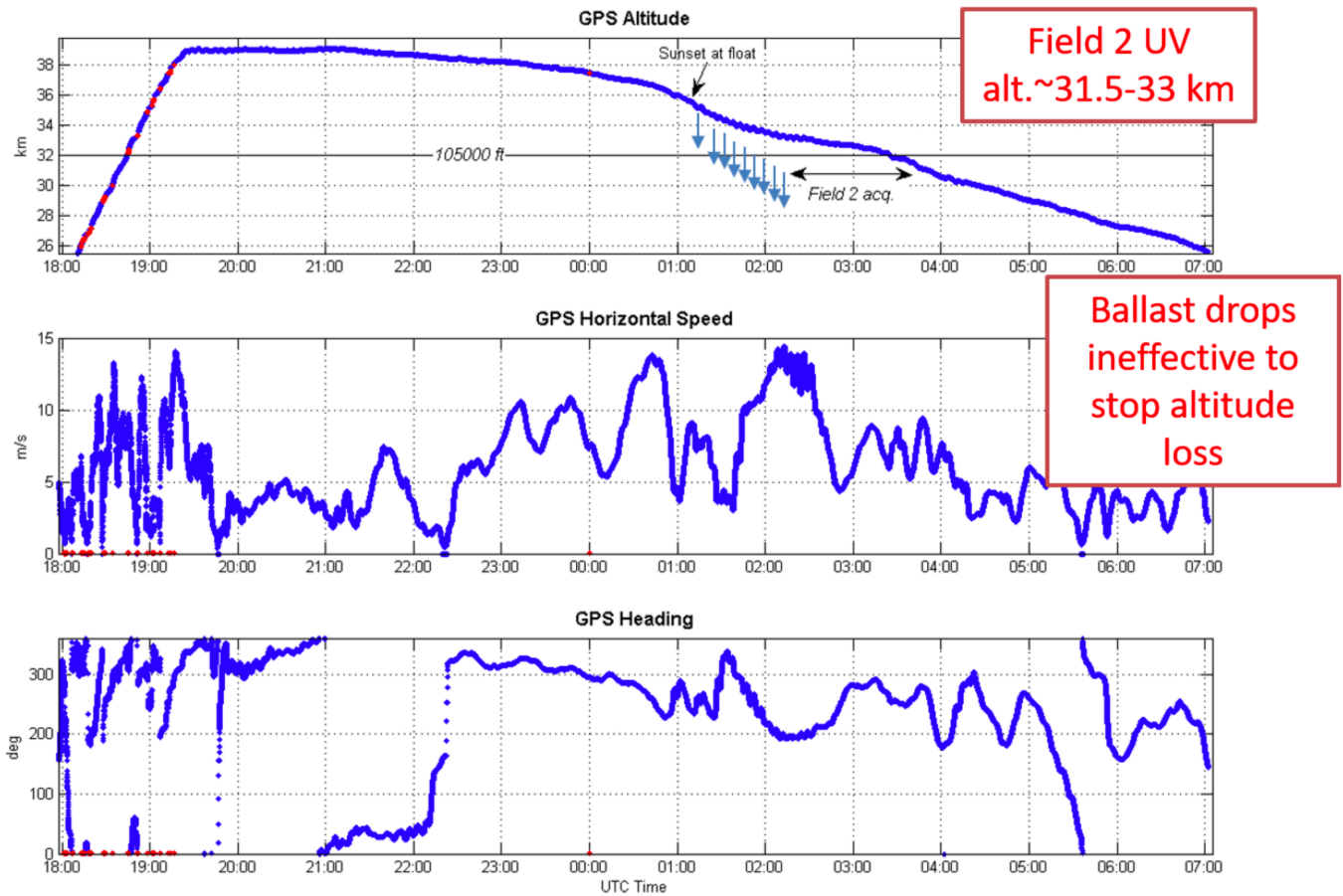}
    \caption{Figure 3 from \citet{2019Montel}. Figure shows the FIREBall-2 gondola altitude, horizontal speed, and heading from the on-board GPS system.} 
    \label{fig:Montel}
\end{figure}

This reduced pointing control was \replaced{a particular issue}{particularly an issue} during the observations of Field 2 (described below in Section \ref{sec:observations}), where the altitude was quickly decreasing and the balloon was rotating through large angles \added{(from 200$^{\circ}$ to nearly 300$^{\circ}$, then back down to 200$^{\circ}$ over a 1.5 hr period)}, requiring a lot of torque for azimuth control. The ballast drops, which also excited a natural vertical (buoyancy) oscillation mode of the gondola, contributed to the large disturbances to position and azimuth during science operations. \added{Figure \ref{fig:Montel}, taken from \citet{2019Montel}, shows GPS measurements of the altitude, horizontal speed, and heading of the payload during the flight. Blue arrows indicate ballast drops, and subsequent oscillations in heading can be seen. Later in the flight, the heading becomes even more unstable, rotating through a full 360$^{\circ}$ before the end of the flight.}

Despite these significant challenges and the extremely degraded observing conditions, the full guidance system performed within the 1" stability requirement throughout the flight. A detailed description of the fine guider and input to the CNES guidance control system is described in Section \ref{sec:guider}.

\subsection{Fine guidance system}\label{sec:guider}

The scientific objectives of FIREBall-2 and the multi-object nature of the spectrograph impose stringent requirements on the instrument pointing. Each science target must be centered in a 80-90 $\mu$m wide slit, corresponding to an accuracy of 6-8 arcseconds.  In addition, FIREBall-2 aims to detect very faint emission, requiring long integration times (up to 100 s). During each integration, the instrument must maintain stability to within a pixel on the detector ($<$1 arcsec) to avoid degrading the image. The purpose of the fine guiding system is to further refine the pre-compensation pointing achieved by the attitude control system (Section \ref{sec:pointing}). The fine guiding system corrects residual errors in the pointing by making on-sky measurements of the instrument pointing offset and providing high-cadence feedback to the attitude control system.    

The entrance to the fine guidance system lies close to the focal plane of the spectrograph where a set of interchangeable masks are mounted on a rotating carousel. Each mask is a duplex system with two layers; an upper mask with a reflective coating directs visible light into the fine guidance system while a lower mask with a pattern of slits transmits UV light through to the detector. Both the upper (guider) mask and lower (science) mask are curved in two dimensions to match the curvature of the focal plane created by the field corrector optics. The focal plane at the mask position is spherical due to the optical design. The guider mask corresponds to an $19.2' \times 16.2'$ patch of sky adjacent to each science field. Masks are selectable and pre-designed for particular target fields. Optical light reflected off the guider mask is fed through a double petzvel lens system into a dedicated guider camera (pco.edge 5.5) with a 2560 $\times$ 2160 pixel CMOS detector. The camera itself is contained within a pressure vessel inside the spectrograph vacuum tank with a window directly above the camera detector in the optical path. A PHOX CameraLink to fiber converter creates an optical fiber output which is then fed through the vacuum tank and out to the guider on board computer. A fiber connection was necessary to maintain coherence between the two CameraLink outputs from the pco.edge without corrupting the signal given the very fast frame rate and length of transmission. An additional PHOX converted the signal back to a CameraLink input into the frame grabber. 

Images of the star field viewed by the guider camera are stored and processed on board by the guider flight computer; these images are also transmitted to the ground via a video downlink. Processing includes identification of stars above a certain SNR threshold and centroiding of their x and y positions in the guider image. Corrections to the instrument pointing are calculated based on the translational and rotational offsets of these stars from their expected positions, producing attitude error signals at a rate of 20-30 Hz. In closed loop, these error signals are fed back into the 4-axis attitude control system, reducing residual errors in the instrument pointing and refining the pointing stability to less than 1 arcsecond. The real-time video feed is used for field recognition and acquisition, target alignment, spectrograph focus, and monitoring the pointing performance during flight. Ground control software for the fine guidance system in communication with the flight computer is capable of controlling the guider camera settings, guide star selection, guide star positioning in the guider field, guider image reduction pipeline, and spectrograph tip-tilt. The same error signals used to refine the guide star positions in fine pointing mode are used to set the initial positions of the guide stars on the guider image. These positions are carefully calibrated pre-flight to ensure that each science target passes through its designated slit on the multi-object slit mask (this calibration is discussed in \citealt{2019Picouet}). 

The optical image that is re-directed to the guider camera is approximately con-focal with the UV image on the detector and the guider camera is used to perform a through focus on each science field in-flight. 

\begin{figure}[]
    \centering
    \includegraphics[width=\columnwidth]{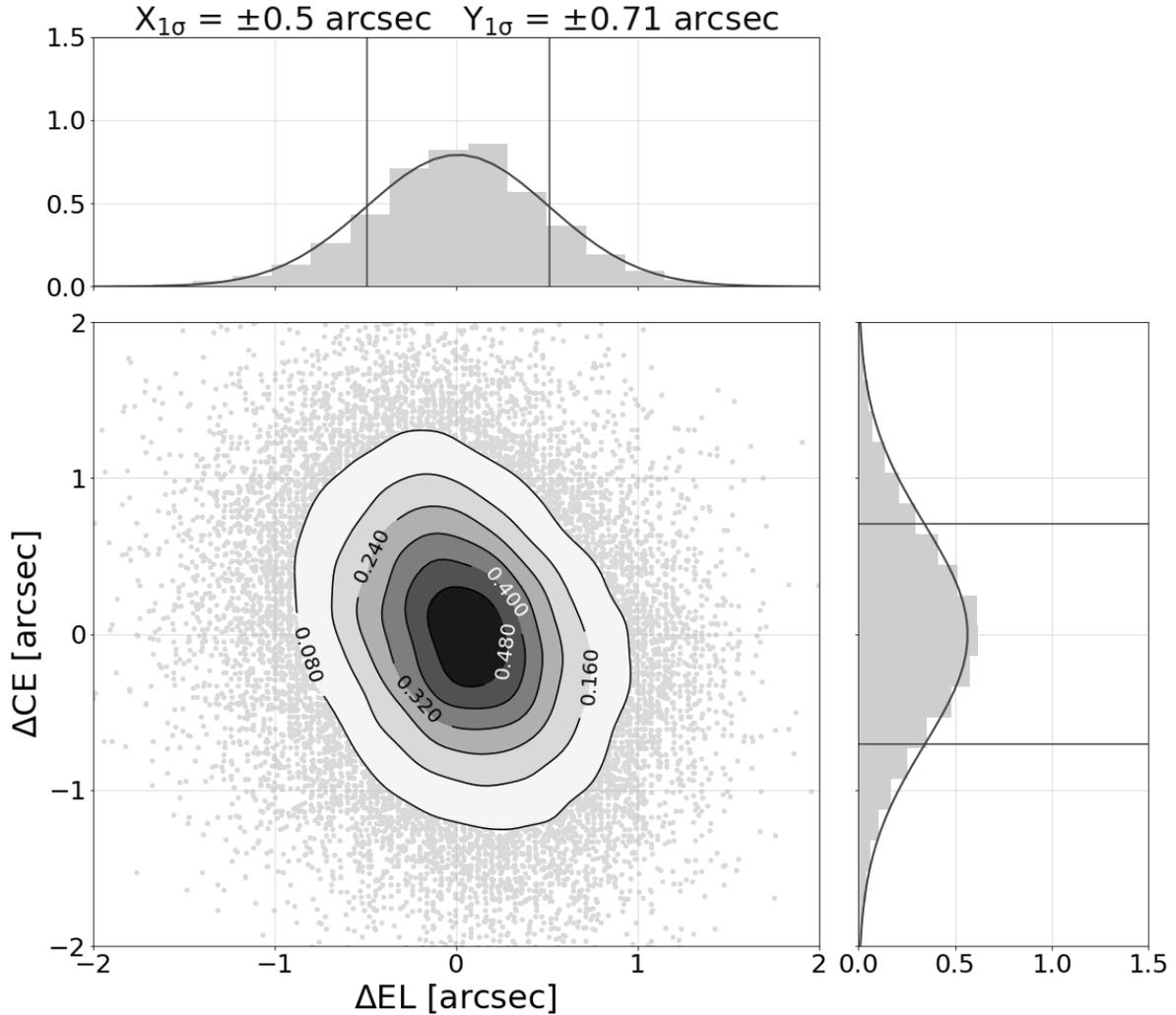}
    \caption{Figure showing pointing performance while observing Field 2 in closed loop fine guiding. The 1D distributions along the top and right panels are probability density histograms fit with a 1D Gaussian. The solid vertical lines are the 1-sigma values for the 1D distribution. The 2D probability density distribution in the center is interpolated with Gaussian kernel density estimation and displayed as shaded contours (where 1-sigma level corresponds to 0.39 $\rm arcsec^{-2}$).} 
    \label{fig:fineguiding}
\end{figure}

The overall performance in fine guiding is shown in Figure \ref{fig:fineguiding}, with excellent RMS over the course of nearly two hours. The balloon descended through nearly 3 vertical km during this observing period, which increases the perturbations on the guidance system. 

\added{A more careful analysis of the guidance control system, the target and guide star selection, and performance can be found in \citet{2019Montel}, \citet{2016Montel}, and in a forthcoming paper, Melso et al. (in prep).}

\subsection{Communications system}\label{sec:comms}

\begin{figure}
    \centering
    \includegraphics[width=0.45\textwidth]{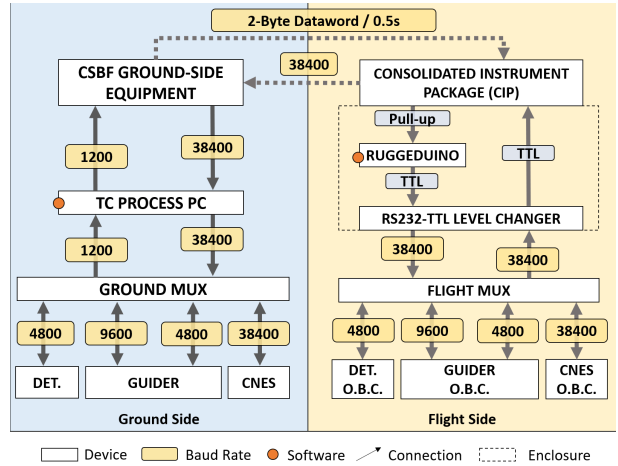}
    \caption{Block Diagram of the communications set-up in 2018. White boxes show devices. Yellow boxes show the baud rates. Orange circle indicates a critical software component. Dashed lines indicate enclosures, not necessarily vacuum sealed.} 
    \label{fig:tmtc}
\end{figure}

The FIREBall-2 communications system consists of three primary transmitters/receivers, all operating via a line-of-sight link: A Consolidated Instrument Package (CIP) uplink/downlink (1200/38,400 baud respectively), a 1 Mbit s$^{-1}$ downlink, and a video downlink. A schematic of the telemetry/telecommand (TM/TC) sub-system of the FIREBall-2 communications system is shown in Figure \ref{fig:tmtc}. This was by far the most complex part of the communications system, and required both interleaving of commands from four different inputs, and a downgrade from 38,400 baud to 1200 baud for uplinking commands. Due to overheads on the uplink per dataword the speed was effectively 32 baud.

The CIP uplink/downlink provided by CSBF took an input from a multiplexer (DBC, Inc., model SR-04) which slowed down and interleaved commands from the detector ground computer, fine guider ground computer, and gondola control ground computer to match the uplink speed. A complementary multiplexer on board de-multiplexed the transmitted signals and relayed them to the appropriate on board computer. Because of the slow uplink speeds, commanding from the ground was laborious and nearly all typical flight actions were scripted ahead of time. The downlink followed a similar path, where the on board computers (detector, guider, and gondola OBCs) sent signals to the multiplexer which sent them down through the CIP. Because the downlink was significantly faster than the uplink housekeeping data for the Gondola OBC and the guider computer were sent via the CIP. Intermittent housekeeping data from the detector computer was also transmitted during flight but the images generated by the detector and associated housekeeping used the faster 1 Mbit s$^{-1}$ downlink. 

The 1 Mbit s$^{-1}$ downlink was used to send cropped images from the science detector. This was necessary both to verify that the targets were falling into the slits on the mask and to allow for nearly real time image processing. As images were generated by the detector, the 1 Mbit s$^{-1}$ downlink would grab the most recent one, strip the overscan and prescan regions, compress it with the most recent housekeeping file, and relay the package to the CSBF transmitter. The time to send a full frame image through was roughly 80 seconds, which necessitated not downlinking every image. Roughly 50\% percent of images were transmitted during flight, with the remaining images collected after the flight including over-scan and pre-scan regions of the downlinked images. \added{The downlink was stable during the flight. Only one image out of several hundred was corrupted during the downlink process. This stability was a result of CSBF's very reliable communication equipment and significant ground testing to eliminate dropped bits as much as possible. One unexpected issue during flight was a strain on the detector computer, which also managed the downlink. The processing power required for encoding images and sending to the 1 Mbit s$^{-1}$ downlink while also operating the \nuvu\ controller would freeze the flight computer and require a restart. This freezing was correlated with the rate of image acquisition. A simple re-start would correct the issue, but image acquisition was stopped while this happened. For future flights, the detector computer has a faster processor which should eliminate this problem.}

The video downlink consisted of a VGA to video adapter for the guider on board computer that would process the video generated by the guider camera. This was sent to a transmitter and the downlink was then relayed to a television screen on the ground and was used as described in Section \ref{sec:guider}. 

\section{Flight on September 22nd, 2018}\label{sec:Flight}

FIREBAll-2 was launched from a 40 MCF balloon from Fort Sumner, NM on the morning of September 22nd, 2018 at 10:20 AM MDT. At launch, all flight systems were nominal and the payload experienced minimal jerks or acceleration during the launch. Throughout the ascent phase (from the ground to float altitude of 128,000 ft) the payload systems and communication continued to behave normally. The payload reached float altitude at approximately 1 PM local time, northeast of the launch site.

\begin{figure}
    \centering
    \includegraphics[width=0.45\textwidth]{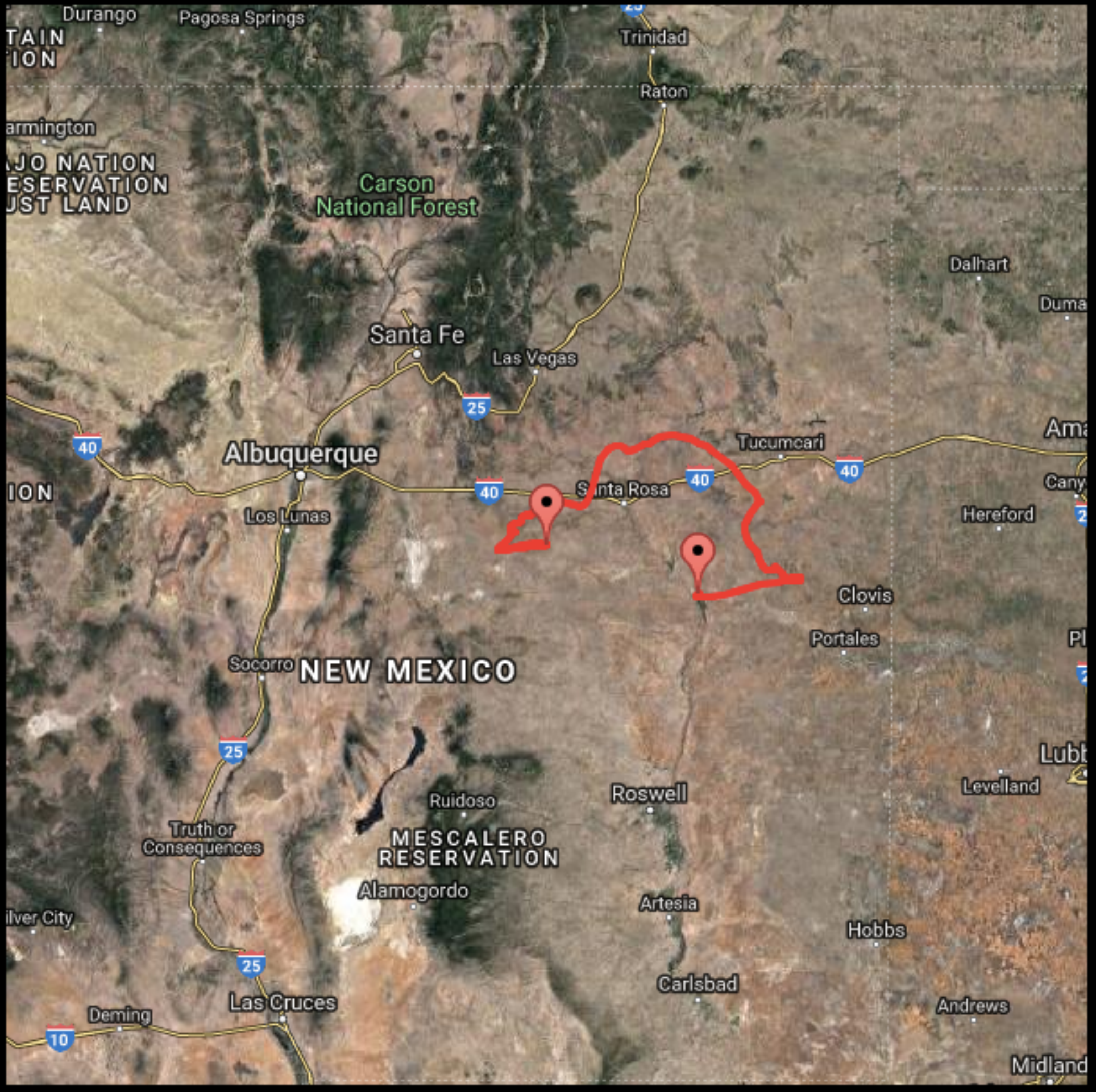}
    \caption{CSBF generated map of trajectory for FIREBall-2 flight on September 22nd, 2018. The flight duration was approximately 13 hours at float after a 3 hr ascent.} 
    \label{fig:flightmap}
\end{figure}

\begin{figure}[]
    \centering
    \includegraphics[width=0.45\textwidth]{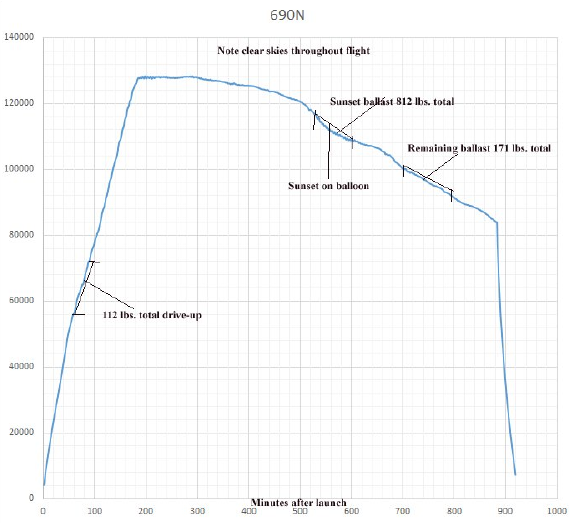}
    \caption{CSBF generated plot showing balloon and payload altitude vs. time since launch. Indicated are sunset on the balloon and several ballast drops.} 
    \label{fig:flightprofile}
\end{figure}

The stratospheric winds were very low during the flight and so the balloon traveled slightly north towards Santa Rosa, NM, before eventually drifting south west and ended up over Vaughn, NM. The GPS map of the flight is shown in Figure \ref{fig:flightmap}. The altitude of the payload over the course of the flight is shown in Figure \ref{fig:flightprofile} including ballast drops and other events. The balloon is expected to lose altitude as the sun sets due to changing thermal conditions and then stabilize at a lower altitude after sunset. Ballast is dropped during sunset to minimize the altitude loss, and the expected nighttime altitude was 118 kft to 108 kft, depending on the particulars of the atmosphere and weather. In part due to anomalies discussed below (Section \ref{sec:balloon}), the balloon lost altitude starting mid-afternoon and continued to descend through the night. The most likely cause of this was a hole allowing helium gas to escape. Ballast drops did not help to slow the descent and additionally created some extra pointing challenges since the timing excited a 5 minute resonant frequency in the balloon altitude. The balloon altitude was below the required science minimum altitude of 32 km at 660 minutes after launch (roughly 9:20 PM MDT). The team continued to collect data throughout the remainder of the flight until it was required to power down all systems by CSBF prior to termination. The balloon and payload were separated at roughly 1 AM MST, taking about an hour to land. 

The flight was terminated 50 miles west of Fort Sumner, near the municipality of Vaugn, NM. The payload was recovered on September 23rd. The landing, due in part to higher winds at night, was rough and both large optics sustained edge fractures. The siderostat fracture was minimal, while the primary mirror lost roughly 5\% of the area including the location of two of the six bond pads, which hold the optic to its mount. The damage for both mirrors was on the same side so was likely a result of a hard landing. In addition, the large mirror of the field corrector was cracked at the edge of the mirror along one of the bond locations. Additional inspection of the remaining mirrors had been conducted. A measurement of the surface figure of both large optics indicated that despite the damage the surface shapes were not changed significantly and there are no additional cracks or fractures that could propagate through the optic. Both optics have since had the broken edges sanded down and were re-aluminized at Goddard Space Flight Center in 2019.

\subsection{Flight anomalies}\label{sec:balloon}

During the flight, there were three anomalies, two related to the balloon. During the launch, the spool that keeps the balloon in position during inflation flew off the inflation vehicle. A subsequent investigation by CSBF into the origins of this is ongoing. Both CSBF and NASA believe this was unrelated to the later anomaly.

After 3 hours at float altitude the balloon began to lose altitude, likely due to a hole that was either present from the launch or developed while at float. As a result, the payload was above the science minimum altitude of 32 km for only $\sim$1 hour after astronomical twilight on the balloon. The atmospheric transmission in the FIREBall-2 bandpass is highly dependent on atmosphere, with a reduction of 10\% for every 3 km of altitude lost. As the flight went on the UV throughput dropped precipitously, which severely impacted our ability to observe the faint targets we had selected.

In addition to degrading the overall throughput the hole in the balloon caused the balloon to deviate from the expected sphere to a more teardrop shape. This teardrop served to direct and focus reflected moonlight into the telescope and spectrograph tank. The scattered light consisted primarily of visible light and was able to impinge on the detector via a small un-baffled direct path from the spectrograph tank top to the back surface of the detector PCB mount, which was then able to reflect onto the detector surface. This possible trajectory for light was not accounted for in the optical analysis. Since the recovery of the payload we have attempted to fully recreate the signature of the scattered light to allow us to determine both the most likely optical path of the scattered light and develop mitigation strategies for future flights.

The altitude loss had a follow-on effect of causing the balloon motion to be highly disturbed. As the balloon and payload descended the wind direction shifted significantly and caused increased noise in the pointing system, degrading performance as described in Section \ref{sec:pointing}. Additionally, the ballast drops caused the balloon to rise and fall in the atmosphere on about a 5 minute cycle. All of these motions were not expected --- typical balloon motion at altitude is very stable, predictable, and tranquil. The fact that the pointing performance was well within spec despite these significant challenges is a testament to both the quality of the pointing system design, implementation, and the team.

The third anomaly involved the Newport RV350 series rotation stage, which is used to keep the spectrograph tank tracking with the rotation of the earth. The stage stopped responding to commands at 10:33 PM local time, or 2.5 hours before the end of the flight. At the time the hole in the balloon and the low altitude of the payload meant there was minimal impact to the science performance. Had the balloon had a nominal flight performance, however, this failure would have been a significant problem and resulted in a loss of more than half of the observing time. Upon inspection post-flight, it was determined that an elastic ring at the end of the motor had a screw that was too short and it did not have enough thread length to hold indefinitely, and under torque. During the flight, while attempting to rotate, the threads of the screw gave way. The broken bolt has since been replaced by a longer one and the stage performed normally.

\subsection{Field Selection and Observation}\label{sec:observations}

Four fields were selected for the 2018 Fort Sumner campaign. Observations with FIREBall-2 target dense galaxy fields at z=0.7 tracing known large scale structure. With the initial launch planned in mid-September from Ft. Sumner NM, several survey fields were visible during the launch window and were also within the 40-60 deg elevation limitation due to balloon and gondola obstruction. Several surveys were examined in detail including BOSS, WIGGLEZ, CANDLES, UKIDSS, CFHTLS, VIPERS, and DEEP2. DEEP2 is a galaxy redshift survey of $\sim$53,000 galaxies conducted from 2002-2005 that used the DEIMOS spectrograph on Keck II \citep{2013Newman}. DEEP2 covered four fields, each roughly 120' by 30' in area, and included the Groth survey strip, a zone of very low extinction, and two SDSS deep strips. The fields were sorted by redshift ranges z = 0.636 - 0.751, z = 0.975 - 1.025 and z = 0.275 - 0.325 for Ly-$\alpha$, OVI and CIV respectively. For Ly-$\alpha$ targets, redshifts z = 0.682 - 0.696 were eliminated since the emission line would coincide with a bright sky line.

Aside from scientific considerations, several observational constraints needed to be considered when selecting the fields that FIREBall-2 targeted. The potential dates for the Fort Sumner launch window, the single night of observation, and the restriction on elevation range due to the balloon and gondola provided limits on potential fields. In addition, selected fields needed an adequate number of stars at the desired bands and magnitudes that fall in the field of view of the guidance system and act as guide stars.

\subsubsection{Targets within fields}

After fields were selected, the slits needed to be filled with desired targets. The criteria for selection included maximizing the number of Ly-$\alpha$ galaxies at z=0.7, spectral lengths of up to a few arcmin per galaxy, providing sufficient space between spectra to avoid overlap on the detector and avoiding redshifts where important galaxy emission lines would overlap on atmospheric features (NO in the upper atmosphere in particular at 2030 and 2047 \AA). When additional space was available that couldn't be filled by a Ly-$\alpha$ target, OVI and CIV galaxies were also used, in that order. The slit length in the cut masks was slightly larger than planned due to the strength of the laser, so in practice there was some slit overlap.

For the mid-September launch from Ft. Sumner, NM, the DEEP2 Fields 2, 3, and 4 were chosen and several mask alternatives were produced and reviewed. Field 2 contains 190 Ly-$\alpha$ galaxies, 372 OVI galaxies, and 2 CIV galaxies. The mask areas in Field 2 were chosen by eye to contain a maximized number of Ly-$\alpha$ galaxies at z = 0.7. 

The star fields were chosen from SDSS Sky Server D12 with magnitudes between 12.0 and 18.0 in the g and r bands. The stars and exact mask coverage areas are chosen to ensure an adequate number of guide stars exist for each target field. FIREBall-2 has resolution on the scale of $\mu$m; therefore, distortions from coordinate conversion and telescope properties have to be accounted for and corrected. 

Due to the low altitude, only FB Field 2, a sub-region from the DEEP2 survey field \citep{2013Newman}, was observed for a significant length of time. Field 2 was acquired at 7:30 PM local time, prior to sunset on the balloon, which is 30 minutes after sunset on the ground. The field was lined up using a standardized acquisition procedure based on the known positions of guide stars until science operations began after the sunset ballast drops. FB-2 observed Field 2 until 9:37 PM MDT for a total of about 45 minutes of exposure time post twilight.

After leaving Field 2 we moved to the center of M31 and then a QSO field (center at RA, Dec: 17.87698, 34.563395) near QSO SDSS J011133.38+343028.5. These target changes were an attempt to observe brighter objects as the payload lost altitude. While data was acquired at each of these locations the altitude was less than 32 km and atmospheric transmission in the balloon window was significantly degraded. 

\subsection{In flight performance}\label{sec:performance}

The instrument in-flight performance is described in more detail in \citet{2019Picouet} but briefly summarized here. 

Post flight processing of the data from Field 2 was used to determine the in-flight performance of the spectrograph. Due to the degraded conditions described in Sections \ref{sec:balloon} and \ref{sec:observations} the performance of some systems were sub-optimal.

Several objects are detected in the co-added data from Field 2. The brightest was continuum emission from a UV bright horizontal branch star with $M_{FUV}$=17.8, which was detected by FB-2 with SNR$>$15 (ID: 2MASS J16515894+3459322, known as Bright Star 1 hereafter). Two fainter UV bright stars were also detected. Bright Star 1 was used to measure the system efficiency, finding 30\% degradation compared to the ground calibrations with atmospheric losses. This is likely due to a combination of uncertainties including target centering, true atmospheric transmission, focus, etc.

In flight spatial resolution is calculated as full width at half max (FWHM) 7", 40\% greater than the 5" expected based on the optical design after convolution with the slit widths and performance on the ground. Part of this degradation may be due to defocus --- a 50 $\mu$m de-focus in flight could lead to a 1 arcsec deterioration in spatial resolution along the slit. Temperature variations in components of the spectrograph and tip/tilt stage can also change the focus. 

In flight spectral resolution could not be measured because the UV bright stars that we observed have no spectral features in this narrow bandpass. There are some features from atmospheric transmission on the UV bright star that can be used, but otherwise we rely on ground calibration. In earlier tests we found a spectral resolution of R$\sim$1600 for a diffuse object (in part due to the wide slits), R$\sim$3500-4000 for a point source \citep{2019Picouet}.

The best PSF shape on the ground is 20\% larger than expected (5.5" instead of 4.5"). The spectrograph was designed for 3" PSF with an additional $\sim$1" degradation because of lowered surface quality of FC2 during polishing. The PSF degradation beyond that may be due to spherical aberrations introduced by the grating but this is pending verification.

\section{Preliminary data analysis}\label{sec:data}

\added{While the flight provided validation of a number of technology advances, the science} data was significantly degraded due to the flight anomalies. Here we discuss the various noise sources and their impact on the data. A detailed discussion of the detector performance can be found in \citet{2019Kyne}, which also includes a significant discussion on the performance of Teledyne-e2v's CCD201-20s in a lab setting.

We are taking steps for the next flight to reduce the effect of the noise sources described below. This includes exploring slowing down the clocking speed to 1 MHz, making adjustments to minimize other sources of noise such as cable length, serial clocking scheme, and increasing the operating temperature of the CCD for better charge transfer efficiency (which involves a trade off with increased dark current). Reducing the impact of smearing will also reduce the impact of cosmic rays on the images as their tails will inevitably be shorter.

\subsection{Detector in-flight performance and measured noise}

In typical visible wavelength observations the largest source of noise is sky background. In the UV the sky background is orders of magnitude lower than in the visible \citep{1998Leinert}, typically leaving detector noise as the limiting noise contribution. For an EMCCD multiple sources of detector noise need to be considered. These include clock-induced-charge (CIC), read noise (RN), dark current, multiplication gain noise, and any background light sources (scattered light, sky noise, etc). In this case, the primary contribution of noise in the data was the excess scattered light from the balloon onto the detector surface via a parasitic optical path that was not baffled. This contributed to a varying background on the order of 0.7 e- pix$^{-1}$ frame$^{-1}$, which is well above the count rate limit of 0.1 e- pix$^{-1}$ frame$^{-1}$ for photon counting \citep{2019Picouet}. Our analysis of the resulting data is focused on understanding and reducing the effect of this scattered light as much as possible. Here we discuss the scattered light and additional sources of noise in the detector system.

\subsubsection{Cosmic Ray Rate}

\begin{figure*}[t]
    \includegraphics[width=\textwidth]{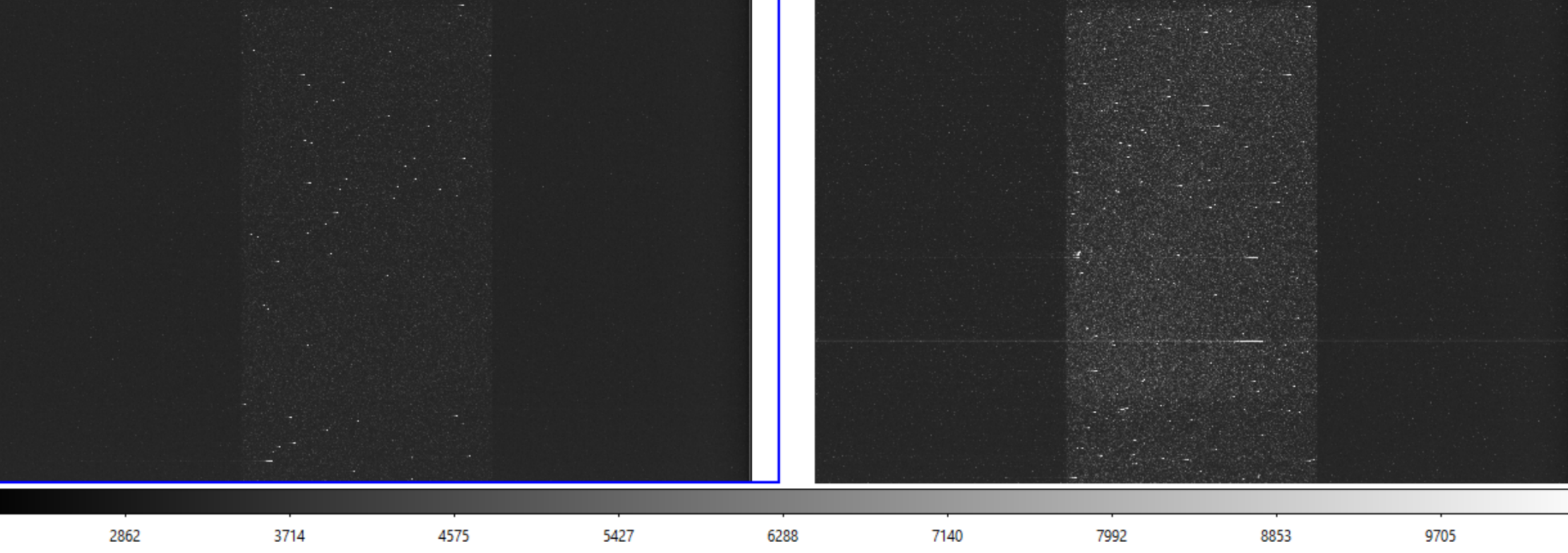}
    \caption{Figure from \citet{2019Kyne}: Detector cosmic ray rate measured during the 2018 flight in Fort Sumner, NM. \textbf{Left:} 30 second exposure. \textbf{Right}: 50 second exposure with a particularly large cosmic ray hit with significant smearing into the overscan region. Overall, we measure a cosmic ray rate of 5 - 7 cosmic rays per second at 128 kft.\\}
    \label{fig:flight_cr_rate}
\end{figure*}

One significant consideration in operating an EMCCD, especially at balloon altitudes, is the effect and impact of cosmic rays on the data. Pixels with cosmic ray hits in an EMCCD will be amplified in the gain register just like any other pixel and have a tendency to then spill over into the trailing pixels (due in part to the event smearing described in Section \ref{sec:smear}). While this issue has been largely mitigated in more modern EMCCD architectures (those being tested for WFIRST), the FIREBall-2 devices do not have the overspill register implemented. During the flight we measure 5-10 cosmic rays per second, contaminating 300-600 pixels (10-15\% of the image on average and up to 25\% in the most extreme cases) depending on the exposure time with longer exposures more affected. Figure \ref{fig:flight_cr_rate} shows example images with the full extent of cosmic ray smearing.

\subsubsection{Dark current and other noise}

Detector noise performance was difficult to disentangle from the effect of excess scattered moonlight from the balloon. With the doors of the gondola closed and tank shutter closed we measured a noise of 1.7$\times$10$^{-4}$ e- pix$^{-1}$ s$^{-1}$. This is likely a combination of dark current, light leaks due to an imperfect shutter/tank cap design, and Cherenkov radiation, and was 10 times higher in-flight than during ground testing. The read noise of the entire system was around 90e- with a pre-amp gain of 0.53 ADU e-$^{-1}$. We operated in a range of gain modes with amplified gain of between 400 and 2000 e-/e- depending on the circumstances. These gain values are typically lower than the normal high values used in photon counting mode because of the excess scattered light.

\subsubsection{Scattered light and residuals}

\begin{figure*}[bt!]
    \centering
    \includegraphics[width=0.95\textwidth]{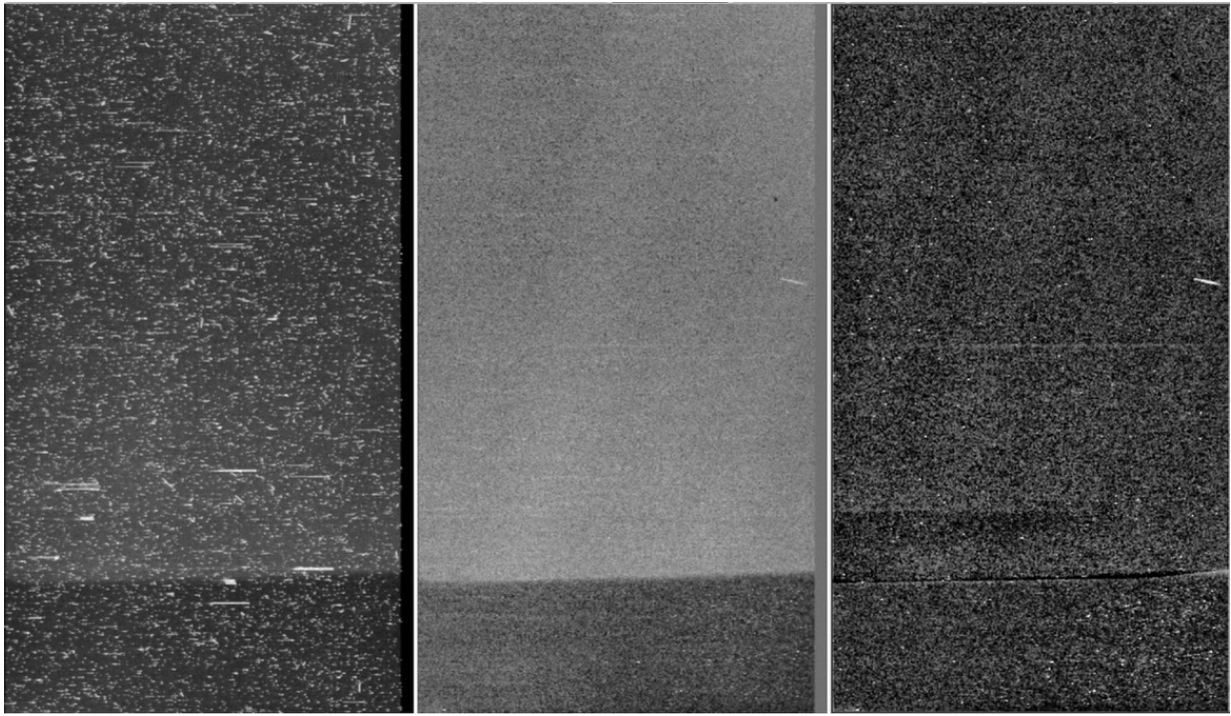}
    \caption{Field 2 summed images. \textbf{Left}: The summed image before cosmic ray removal. \textbf{Center:} The summed image after de-smearing and cosmic ray removal. \textbf{Right:} The summed image after de-smearing, cosmic ray removal, and subtracting the background scattered light. The sharp drop-off in scattered light on the bottom fifth of the image makes the subtraction difficult, yielding a residual along the edge of the scattered region. The dark region at the top of the detector was not delta-doped due to the placement of a mask during the MBE process. These pixels are not responsive to UV light. The bright slash of pixels in the middle right of each image is a cluster of hot pixels. \added{Additional hot pixels are clustered in the bottom right of the images. This detector had multiple defects but the highest QE out of the available devices, which is why it was selected.} All images have been smoothed with a Gaussian kernel with sigma of 1.5 and a radius of 3 pixels. The emission from the brightest star, Bright Star 1, is visible in all three images as a horizontal line roughly in the middle. Emission from other sources can be seen faintly in the right most image.} 
    \label{fig:summed}
\end{figure*}

The scattered light was primarily moonlight being scattered and re-directed from the deflated balloon. During the flight the moon was 93\% full. Most of our target fields were more than 45$^{\circ}$ away from the moon and we anticipated that moonlight wouldn't directly enter the optical system. Previous tests observing Vega indicated that the expected scattered light within the optical path would be on the order of $\leq$10$^{-4}$ of the in-band signal. The deflated balloon shape, however, acted as a lens and caused illumination that bypassed the optical path entirely and instead had an un-baffled view of the detector board and nearby optics. 

The bulk pattern of scattered light (shown in the central panel of Figure \ref{fig:summed}) fluctuated on 5 minute intervals, in sync with the balloon vertical oscillations. We were able to re-create much of the signature of the scattered light after the fact using pseudo-dome flats and other methods to fully illuminate the entrance aperture to the spectrograph tank, beyond just the pupil. 

\begin{figure}[]
    \centering
    \includegraphics[width=0.45\textwidth]{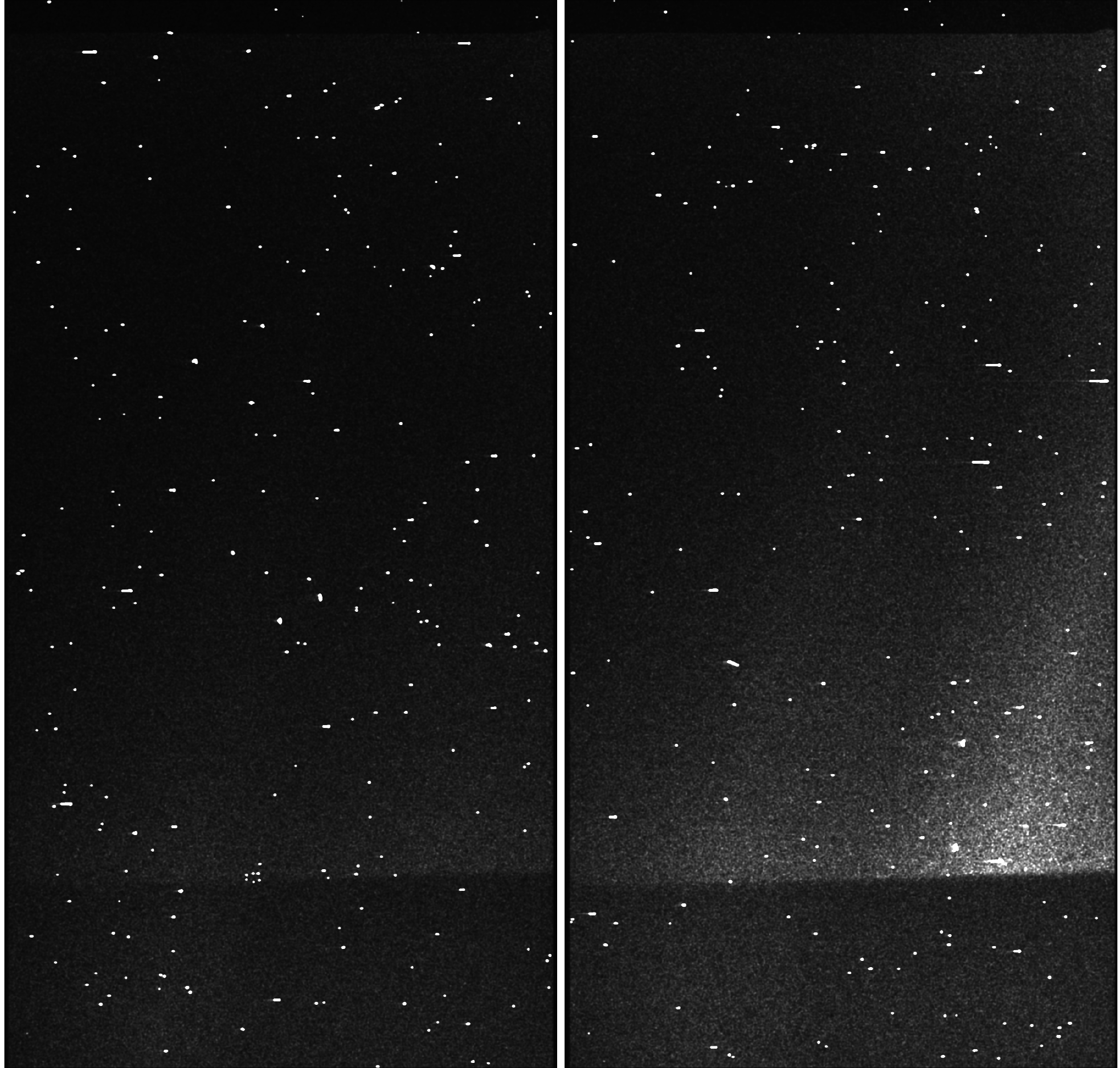}
    \caption{Two sequential images showing the brightness of the `flashes' visible in the data. The scale and color bars are the same for both images. White dots are uncorrected cosmic rays. \textbf{Left:} a typical frame with normal scattered light pattern. \textbf{Right:} the frame immediately proceeding which contains a `flash'.} 
    \label{fig:flashes}
\end{figure}

In addition to large scale scattered light there were occasional flashes that are visible in the flight data, shown in Figure \ref{fig:flashes}. These flashes are typically much brighter ($\sim$2.5 times) than the background level of scatter and have a characteristic spatial signature. We have also been able to re-create them in the lab \added{post-flight} and believe, due in part to their transient nature, brightness, and distinctive presentation, that they are the result of light reflecting off of a metal surface on the detector PCB mount. 

\added{We are currently running a comprehensive scattering analysis of the as-built 2018 spectrograph and gondola using the FRED software package. This analysis should identify the exact optical paths that created the both the bulk scattered light and the flashes and confirm our post-flight experiments.}

For a future flight, we will implement a multi-pronged mitigation strategy to reduce the effect of scattered light. First, we plan to reduce the field of regard of the balloon from the spectrograph tank opening by the addition of baffling that will stretch across the gondola doors when open. This limits the upper altitude accessible by the siderostat and so will be carefully positioned to minimize the impact on the science target observability. The second step is to position baffling at the top of the spectrograph tank, again to reduce lines of sight from beyond the parabola into the spectrograph. The third step is to increase the level of baffling within the spectrograph tank itself. This includes more extensive baffling around and beyond the pupil, above the slits masks, as well as the baffling directly around the detector, to reduce reflective surfaces.

\subsubsection{Event smearing}\label{sec:smear}

The smearing of events in the multiplication register is a common problem in EMCCDs. It derives from a combination of low temperature (which can reduce CTE) and the fast pixel clocking speed (10 MHz in this case). The FIREBall-2 spectrograph set-up required the detector inside the vacuum tank pressure vessel while the \nuvu\ controller box was located outside, connected by an 11 inch cable. This is at the very limit of what \nuvu\ recommends for the 10 MHz pixel clocking speed, and was just barely manageable given the spectrograph configuration. 

The net effect of event smearing is to reduce the measured gain value (since high count pixels will be spread out over 3-4 trailing pixels, lowering the high end of the histogram and increasing the low end). In addition, it makes it difficult to distinguish between events in adjacent pixels vs. a single event that has been smeared. The smearing signature is an exponential and is fairly easy to characterize, but like the amplification process itself, does not necessarily follow the same pattern every time. The impact of smearing is to reduce the number of pixels that are above the standard EMCCD 5$\sigma$ threshold, and a reduction in the apparent gain. A de-smearing algorithm using median absolute deviation is able to reduce the impact of the smear and is detailed in \citet{2019Kyne}.

\subsection{Co-added data analysis}

Here we present the co-added data from Field 2, from the DEEP2 survey. This image was made by co-adding 55 minutes worth of data, taken under sub-optimal conditions as described in Section \ref{sec:balloon}. Figure \ref{fig:summed} shows three different versions of the same field and data. The left most shows the summed image before cosmic ray removal. The center image shows the summed image after de-smearing and cosmic ray removal, with individual images cosmic ray removed prior to summing. The pixels with cosmic rays are filled with data interpolated from the surrounding pixels. The right image shows the summed image after de-smearing, cosmic ray removal, and subtracting the background scattered light. \deleted{The sharp drop-off in scattered light on the bottom fifth of the image makes the subtraction difficult, yielding a residual along the edge of the scattered region. The dark region at the top of the detector was not delta-doped due to the placement of a mask during the MBE process. These pixels are not responsive to UV light.}

The process of co-adding is typical for that used for EMCCDs as described in \citet{2019Kyne}. The multiplication gain is calculated for each individual image in both the image area, pre-scan, and over-scan regions. Smear correction is done using the median absolute deviation (MAD) method and a 6-sigma correction. After the smear correction, the multiplication gain is re-measured. Typically the post-smear correction gain is 2-3 times higher than the pre-smear correction gain, which is expected. Individual images are converted from counts to electrons using a conversion gain measured from a photon transfer curve (1.77 e- ADU$^{-1}$ in this case). Because of the high background, these images were not taken in a pure photon counting mode and instead the analysis uses a gain mode. The EM gain calculated for each image is used to convert the image back to photons, which results is a $\sqrt{2}$ loss of overall efficiency. Individual images are then summed and the total exposure time is used to convert to photons per second.

All three UV bright stars are visible in the summed image, which is significantly contaminated by stray light. A subsequent analysis attempted to remove the large signal from stray light, but no signal from Ly-$\alpha$ was detected around individual galaxies. Summing spectra from many galaxies also did not yield any significant detections.

\subsection{In-flight sensitivity}

Due to the low altitude, scattering from the balloon de-shape, and degraded performance, the overall sensitivity was worse than the FB-1 flight. \citet{2019Picouet} calculates a sensitivity of 80,000 LU (vs. 74,000 LU for FB-1) using the measured emission from the detected UV bright stars. This was significantly higher than the expected sensitivity of 8,000 LU and was driven entirely by increased noise. 

\section{Future flight possibilities}\label{sec:future}

While the payload sustained significant damage upon landing, the resulting analysis has shown that the major components (the two large optics, the gondola structure, the pivot, the spectrograph) can all be re-flown as they are or with limited refurbishment. Additionally, nearly every aspect of the payload and spectrograph system performed as expected and within specifications. We do not need to redesign any subsystems given the excellent performance. Thus, we are working towards a \deleted{2020}\added{2021} re-flight of the FIREBall-2 payload. We will be able to mitigate the effects of scattered light and are investigating some upgrades to the cooling/detector system to reduce the cooling power and heat load required as well as upgrade to the \nuvu\ v3 CCCP controller to reduce the read noise in the detector. We are currently on track for a \deleted{2020}\added{2021} flight attempt from Fort Sumner using similar fields and targets as for the 2018 flight. We are adding an additional field with several QSOs and QSO pairs, in response to the more recent observations of QSO pairs described earlier.

Future upgrades beyond \deleted{2020}\added{2021} (i.e. for a FIREBall-3) may include a re-design of the spectrograph using microshutter arrays and potentially upgraded detector architecture, but are currently in a very preliminary phase.

\section{FIREBall-2 as a pathfinder for future UV telescopes}\label{sec:space}

The FIREBall sub-orbital program has always been carried out with three goals in mind. The first is to train and promote early career instrument scientists as future mission PIs. In this sense, it has already succeeded, generating over 10 PhDs, and giving many students both flight hardware experience and the opportunity to experience a field campaign. Additionally, it provides an opportunity for postdocs to take on important, leadership roles that would otherwise not be accessible to them, giving them training as future PIs.

The second goal has been technology demonstrations. The flight of a UV-optimized delta-doped EMCCD is the first flight test of two truly mission-enabling technologies. These detectors will play key roles in future missions. Additional technology, such as the anamorphic grating, guidance system, and use of a multi-object spectrograph on a balloon payload have also been significant improvements over the existing state of the art. 

Finally, FIREBall has always been a pathfinder for a larger program to observe and map the faint circumgalactic and intergalactic media. While the observations in the visible bandpass have vindicated the concept of directly observing the diffuse gas from the CGM, we have not yet realized the potential of doing these observations in the UV. The \deleted{2020}\added{2021} flight of FIREBall-2 will detect this emission at a redshift of 0.7, but a future mission concept to explore the CGM in the UV from redshifts of z$<$2 is essential in our understanding of galaxy evolution.

While the data gathered from the 2018 flight of FIREBall-2 may not provide a detection of the CGM at z=0.7, the flight was a success in a number of ways. Future flights of FIREBall-2 will improve on the past missions and we remain excited for the future.

\acknowledgments

The authors wish to thank the helpful reviewer for their insightful comments. In addition, we sincerely thank the many and varied supporters over the years of development, design, integration, and flight. Many people have helped out the team in moments of need and we are grateful. We especially thank the many hardware and component vendors who were patient and helpful in all of our questions.

The FIREBall mission is a joint venture of NASA and CNES, and is funded on the US side via the NASA APRA program. The FIREBALL collaboration also acknowledges considerable support from CNES, LAM, and CNRS. The authors wish to thank in particular Mike Garcia, the NASA program scientist for APRA, and Mario Perez, also of NASA, for their continued support for this mission. The Columbia Scientific Ballooning Facility provides the balloon, launch, ground, and flight support for all flights of FIREBall and have been professional, helpful, and informative throughout two flight campaigns in Fort Sumner, NM. In particular, we wish to thank Hugo Franco for his humor and help during 2017 and 2018. We also thank the Balloon Program Office (BPO/NASA) and Debora Fairbrother. 

For excellent advice and assistance on detector testing and optimization, we sincerely thank \nuvu\, especially Olivier Daigle and Yoann Gosselin. We also thank Teledyne-e2v, for their continued support in detector development and testing.

\vspace{5mm}
\facilities{}

\software{}

\bibliography{references}{}
\bibliographystyle{aasjournal}

\end{document}